\documentclass[a4paper,11pt,openbib,final,footheight=-5pt]{scrartcl}
\usepackage[a4paper,includehead,top=20mm,bottom=20mm,right=20mm,left=20mm]{geometry}

\makeatletter
\DeclareOldFontCommand{\bf}{\normalfont\bfseries}{\mathbf}
\DeclareOldFontCommand{\it}{\normalfont\itshape}{\mathit}
\makeatother

\usepackage[utf8]{inputenc}
\usepackage{amsmath}
\usepackage{amssymb}
\usepackage{esvect}
\usepackage{graphicx}
\usepackage{mathtools}
\usepackage{multirow}
\usepackage{rotating}
\usepackage{hhline}
\usepackage{floatrow}
\usepackage{caption}
\usepackage{subcaption}
\usepackage{bm, bbm, upgreek, stmaryrd}
\usepackage[title]{appendix}

\DeclareMathAlphabet{\mathscr}{OT1}{pzc}{m}{it} 
\DeclareMathAlphabet{\mathpzc}{OT1}{pzc}{m}{it}
\newcommand{\m}{^\text{m}}
\newcommand{\M}{^\text{M}}
\newcommand{\p}{\partial}
\newcommand{\free}{\mathpzc{f}}
\renewcommand{\d}{\, \mathrm d }
\newcommand{\ten}[1]{\bm{#1}}

\newcommand{\pd}[2]{\frac{\p #1}{\p #2}}
\newcommand{\begeq}{\begin{equation}\begin{gathered}}
\newcommand{\eqend}{\end{gathered}\end{equation}}
\newcommand{\begal}{\begin{equation}\begin{aligned}}
\newcommand{\alend}{\end{aligned}\end{equation}}
\newcommand{\del}{\updelta }
\newcommand{\Dt}[1]{#1^{\scalebox{0.4}{\textbullet} } }

\newcommand{\eps}{\varepsilon}

\newcommand{\Reff}{\text{ref}}

\newcommand{\cen}{\overset{\text{c}}}

\newcommand{\mi}{^\text{m}}
\newcommand{\ma}{^\text{M}}
\newcommand{\dd}{\, \mathrm d }

\title{\huge Determining parameters in generalized thermomechanics for metamaterials by means of asymptotic homogenization}

\author{Bozo Vazic$^{a}$
\and
Bilen Emek Abali$^{b}$\thanks{Corresponding author: bilenemek@abali.org}
\and
Pania Newell$^{a}$\thanks{Corresponding author: pania.newell@utah.edu}
\and
\\
\small $^a$Department of Mechanical Engineering, \\[-0.1in]
\small The University of Utah, Salt Lake City, Utah 84112, USA
\\[0.02in]
\small $^b$Department of Materials Science and Engineering, Division of Applied Mechanics\\[-0.1in]
\small Uppsala University, Ångströmlab Box 35, 751 03 Uppsala, Sweden
}

\date{} 

\begin{document}
\pagenumbering{arabic}

\maketitle

\begin{abstract}
Advancement in manufacturing methods enable designing so called metamaterials with a tailor-made microstructure. Microstructure affects materials response within a length-scale, where we model this behavior by using the generalized thermomechanics. Strain gradient theory is employed as a higher-order theory with thermodynamics modeled as a first-order theory. Developing multiphysics models for heterogeneous materials is indeed a challenge and even this ``simplest'' model in generalized thermomechanics causes dozens of parameters to be determined. We develop a computational model by using a given microstructure, modeled as a periodic domain, and numerically calculate all parameters by means of asymptotic homogenization. Finite element method (FEM) is employed with the aid of open-source codes (FEniCS). Some example with symmetric and random distribution of voids in a model problem verifies the method and provides an example at which length-scale we need to consider generalized thermoeleasticity in composite materials.
\end{abstract}


\section{Introduction}
The majority of natural (e.g., polycrystals, wood, and bone) and man-made (e.g., fiber-reinforced composites, concrete, ceramics, and metallic foams) materials are heterogeneous at the micrometer ($\upmu$m) length-scale. Heterogeneous materials have differing physical properties within the structure at the so-called microstructure; microstructural, crystal structural, or compositional heterogeneity exists. Heterogeneous materials own their widespread use in engineering and scientific applications (e.g., spaceflight technology, energy conversion, or energy storage) to the combination of inherent or tunable mutually beneficial properties such as low relative density, heat insulation, high heat resistance, and chemical resistance, or extreme hardness, \cite{thompson1982high,torquato2002random,hutmacher2007state,liu2011metamaterials,barchiesi2019mechanical}. Heterogeneous materials, despite being composed of domains possessing distinct physical properties at the microscale, may be modeled accurately at macroscale as homogeneous materials by an effective (homogenized) material-like properties, \cite{kalamkarov2009asymptotic,martinez2017homogenization,rottger2019time}. As expected, physical response on the continuum level is strongly coupled to the microscale heterogeneity for a length-scale near the microscale's length-scale. \\

In the case of thermo-mechanical processes, microscale heterogeneity's role may be significant within a length-scale. For example, due to a mismatch of microscale thermal expansion coefficients, materials subject to high stress or temperature environments (e.g., concrete and bedrock used in nuclear waste storage) exhibit sharp stresses at the macroscopic level \cite{wallner1982thermomechanical}. Hence, damage may occur due to induced thermo-mechanical stresses and minimize the operational lifetime of the component. For this reason, efforts have been made to develop more accurate theoretical models to predict material's physical behavior \cite{germain1973method,maugin2015some,drapaca2019brief}. The evolution of theoretical models necessitates the development of numerical homogenization methods based on averaging different physical fields to obtain the effective physical properties \cite{geers2010multi,matouvs2017review}. The average and the calculation of local field quantities are carried out by solving the underlying physics problem within a so-called representative volume element (RVE) to model the microstructure.\\

In classical Cauchy continuum mechanics, linear elastic models are based on Hooke's law that implies a linear relation between Cauchy stress and strain. In order to encapsulate thermal effects, linear elastic models have been extended to include temperature by adding an extra linear dependency of Cauchy stress, as formulated in Duhamel--Neumann extension in thermomechanics. At a macroscale with several orders of magnitude larger length-scale than the microscale, which is indeed the case in many engineering applications, the aforementioned models are perfectly admissible. However, the same models fail to account for the complex heterogeneous microstructure at the macroscale with a similar length-scale to microscale, accurately \cite{del2014dynamic}. Hence, generalized continuum theories have been developed to counteract the inability of classical continuum mechanics to account for the microstructural effects, \cite{mindlin1965second,maugin2010generalized}.\\

The generalized continuum additionally incorporates higher-order gradients of essential kinematic variables and associated length-scale parameters. The most common application of such models is the strain gradient model, where alongside strain, we have a gradient of strain as an additional state variable \cite{mindlin1965second}. The addition of strain gradient introduces higher-order ``hyper'' stress as a work conjugate of the strain gradient \cite{dell2009generalized}. However, the extension of the strain gradient models to account for temperature poses new challenges \cite{khakalo2019lattice}. One approach follows Coleman--Noll rational thermodynamics \cite{truesdell1984historical} where only temperature is included in Helmholtz free energy in addition to strain and gradient strain \cite{polizzotto2012gradient}. Another approach includes temperature and gradient of temperature into the Helmholtz free energy, \cite{maugin1990infernal,germain1973method, forest1999towards}. Both approaches lack experimental data that would provide additional material parameters arising from the inclusion of temperature and gradient of temperature into the Helmholtz free energy. This problem becomes more challenging for models with both temperature and its gradient as we are adding not one but two additional variables in the free energy formulation. Furthermore, another problem that arises from the addition of temperature gradient is an extra time derivative that appears in the flux term. This choice leads to an extension of Fourier's law into Cattaneo's equation, where alongside conductivity, we have an additional parameter coupled with the time derivative of the temperature gradient \cite{muller2013rational}.\\

Going from the classical continuum to the generalized continuum, homogenization models go from first-order approaches dealing with strain (displacement first derivative) to second-order approaches dealing with strain and gradient of strain (displacement second derivatives)  \cite{fish1997computational,torquato1998effective,HE2020109519}. Particularly, the first-order (to be precise, first-gradient) approach requires strict separation of scales, and adherence to the concept of local action negates the ability to capture microscale geometry and deal with localization problems \cite{ameen2018quantitative}. The second-order approach, by virtue of the generalized continuum, enables us to capture the microscale geometry by introducing length-scale into the material constitutive law \cite{terada2000simulation,geers2001gradient,fish2008mathematical,geers2010multi,misra2021identification}. Although, applications of the first-order homogenization approaches in thermoelastic problems are abundant in the literature, \cite{ozdemir2008computational,dasgupta1994effective, zhang2007thermo}, the second-order homogenization approaches are relatively rare due to previously mentioned problems. Instead of homogenization, multiscale approaches exist, where the finite element method at both scales (FE$^2$) may be used to do a thermo-mechanical analysis of heterogeneous solids \cite{ozdemir2008fe2}. Such an approach requires high-order continuity of macroscale equations, which relies on finite element formulation that should have (at least) $C^1$ regularity in displacement and temperature fields. A significant characteristic of the multiscale asymptotic approach is the ability to avoid continuity requirements owing to the reestablishment of the high-order macroscale derivatives by post-processing. Thus, other researchers \cite{yang2016thermo, fish2019second} used a second-order asymptotic expansion approach to analyze the coupled thermo-mechanical problems. High-order asymptotic models effectively investigate coupled problems by solving periodic functions at the microscale and generating the macroscale displacement and temperature fields \cite{yang2019thermo}. Additional parameters emerge and they need to be explicitly calculated. This work aims to explicitly calculate all the higher-order material terms associated with the generalized continuum model, such as higher-order elastic constants, coupling constants, and parameters associated with temperature. \\ 

In the present study, we only include temperature in our model to avoid extending Fourier's law. In this manner, we analyze the simplest thermo-mechanical model in strain gradient elasticity. We follow the asymptotic homogenization in strain gradient elasticity as introduced in \cite{abali2020additive}, verified in \cite{yang2021verification}, and applied in \cite{vazic2021mechanical}. In order to incorporate temperature in the asymptotic homogenization model, we follow existing methods \cite{terada2010method,forest2000thermoelasticity,forest2001asymptotic}. In doing so, we develop a higher-order asymptotic homogenization model for thermoelastic strain gradient materials that accounts for all of the accompanying higher-order material parameters.\\

The rest of the paper is organized as follows. The higher-order asymptotic homogenization method and computational implementation are explained in detail in the second section. Numerical results and a discussion of higher-order parameters are presented in the third section, followed by the conclusion. 

\section{Methodology}
We follow the asymptotic homogenization method \cite{pinho2009asymptotic,abali2020additive} and extend it to thermomechanics. The microstructure is denoted by $\ten y$ and in the rest of the paper we will call it microscale; and its corresponding homogenized continuum is denoted by $\ten X$, called macroscale. Their transformation is handled by a so-called homothetic ratio, $\epsilon$. Thus, we circumvent a scale separation which enables us to use the same coordinate system for both length-scales. The approach is based on the ``known'' microscale leading to the ``sought after'' parameters at the macroscale.\\

We begin with thermomechanics at microscale and use balance of momentum for calculating the displacement, $\ten u$, by a defined stress, $\ten\sigma$, under a given (specific) body force, $\ten g$, as follows:
\begeq\label{balanceOfM}
\rho\mi \ddot{u}\mi_i - \sigma\mi_{ji,j} - \rho\mi g_i = 0 \ ,
\eqend
where we use a comma notation denoting the space derivative and $\rho\m$ is the (known) microscale mass density. Herein and henceforth, we use standard continuum mechanics formulation with summation convention over repeated indices. Similarly, the balance of internal energy reads
\begeq\label{balanceOfE}
\rho\mi \dot{u}\mi + q\mi_{i,i} - \rho\mi r = \sigma\mi_{ji} (\dot{\eps}\mi_{ij}) \ ,
\eqend
where the specific (per mass) internal energy, $u$, and heat flux, $\ten q$, need to be defined. Supply term, $r$, is the specified internal thermal source. It should be noted that strain used in Eq.\,\eqref{balanceOfE} is defined in linear form shown below,
\begeq\label{lineStrain}
\eps\mi_{ij} = \frac12\big( u\mi_{i,j} + u\mi_{j,i} \big) \ .
\eqend 
Geometric nonlinearities are ignored such that the reference frame is equal as the current frame. Therefore, the rate is simply the partial time derivative in the reference frame that we choose as the known initial (undeformed) configuration. Generalization to higher order is adequate by using an energy formulation. By choosing the specific Helmholtz free energy:
\begeq\label{helmholtz}
\free \mi = u \mi - T \mi \eta \mi \ ,
\eqend
where $T\mi$ is the microscale temperature, and $\eta\mi$ is the microscale specific entropy. Here we introduce the first simplification, $\free\mi=\free\mi(T\mi, \ten\eps\mi)$, indicating that the free energy depends only on temperature and strain. This approach is valid in thermoelasticiy and we circumvent ourselves from justifications like objectivity (usually done in rational thermodynamics) and use a more direct approach of defining the free energy in an axiomatic manner (as in continuum thermodynamics or in non-equilibrium thermodynamics) where the internal energy is simply defined. By inserting $\dot{u}\mi = \Dt{(\free\mi+T\mi\eta\mi)}$ into the Eq.\,\eqref{balanceOfE}, dividing by $T\mi$, using 
\begeq\label{entropy}
\eta\mi = -\pd{\free\mi}{T\mi} \, , \
\sigma\mi_{ji} = \rho\mi \pd{\free\mi}{\eps\mi_{ij}} \ ,
\eqend
and since there is no dissipative stress in the system (elasticity), we obtain 
\begeq
\rho\mi \Dt{(\free\mi+T\mi\eta\mi)}+ q\mi_{i,i} - \rho\mi r = \sigma\mi_{ji} (\dot{\eps}\mi_{ij}) 
\ , \\
\rho\mi \Big( \pd{\free\mi}{T\mi} (\dot{T}\mi) +  \pd{\free\mi}{\eps\mi_{ij}} (\dot{\eps}\mi_{ij}) 
+ (\dot{T}\mi)\eta\mi + T\mi(\dot{\eta}\mi) \Big)
+ q\mi_{i,i} - \rho\mi r = \sigma\mi_{ji} (\dot{\eps}\mi_{ij}) 
\ , \\ 
\rho\mi (\dot{\eta}\mi) + \frac{1}{T\mi} q\mi_{i,i} - \rho\mi \frac{r}{T\mi} = 0 \ .
\eqend
After rewriting the heat flux in a straight-forward manner, we obtain the balance of entropy:
\begeq\label{balanceEntropy}
\rho\mi (\dot{\eta}\mi) + \Big( \frac{q\mi_i}{T\mi} \Big)_{,i} - \rho\mi \frac{r}{T\mi} = -\frac{q_i}{(T\mi)^2} T\mi_{,i} \ .
\eqend
The right-hand side is the entropy production in thermoelasticity which is positive according to the second law of thermodynamics. This assertion results in a restriction for the heat flux, herein, we use a linear relation called Fourier's law:
\begeq\label{fourier}
q_i = \kappa\mi_{ij} T\mi_{,j} \ .
\eqend
where $\kappa_{ij}\m$ is the thermal conductivity. Furthermore, since $\free\mi=\free\mi(T\mi, \ten\eps\mi)$, we have $\eta\mi=\eta\mi(T\mi, \ten\eps\mi)$ as a simple mathematical fact based on Eq.\,\eqref{entropy}---often it is introduced as a principle of equipresence, but there is no need for such a principle, since there is a mathematical justification for this, we refer the readers to \cite{014} for further details. By summing up the equations for thermoelasticity, we obtain
\begeq\label{balancEntropy1}
\rho\mi (\ddot{u}\mi_i) -  \bigg( \rho\mi \pd{\free\mi}{\eps_{ij}} \bigg)_{,j} - \rho\mi g_i = 0 \ , \\
\rho\mi \Dt{ \bigg( \pd{\free\mi}{T\mi} \bigg)} - \frac1{T\mi} q\mi_{i,i} + \rho\mi \frac{r}{T\mi} = 0 \ .
\eqend
In this manner, the whole formulation is reduced to one scalar function, Helmholtz free energy and its definition. Corresponding to the linear material model (Fourier's law) used in heat flux, we continue to model the microscale as a linear thermoelastic material. Thus, we use linear elastic model with the known stiffness tensor, $\ten C\mi$, thermoelastic interaction, $\beta\mi_{ij} = C\mi_{ijkl} \alpha\mi_{kl}$, with the well-established coefficient of thermal expansion, $\ten\alpha\mi$. In this setting, the Helmholtz free energy is modeled as a quadratic one,
\begeq\label{defEng}
\free\mi = -c\mi T \bigg( \ln\Big(\frac{T}{T_\Reff}\Big)  - 1 \bigg) 
+  \frac1{2\rho\mi} \eps\mi_{ij} C\mi_{ijkl} \eps\mi_{kl} 
+ \frac{(T-T_\Reff)}{\rho\mi} \beta\mi_{ij} \eps\mi_{ij} \ .
\eqend 
To compare microscale and macroscale Helmholtz free energies, we simplify Eq.\,\eqref{defEng} by expanding logarithmic temperature function through Taylor expansion as:
\begeq\label{tylor}
\ln \xi = \frac{\xi-1}{\xi} + \frac{(\xi-1)^2}{2\xi^2} + ...\quad \xi \geq \frac{1}{2}
\eqend
where $\xi = T/T_\Reff$. Thus, first expression on the right hand side of Eq.\,\eqref{defEng} is expanded,
\begal\label{tylorExpRHS}
\ln \Big(\frac{T\mi}{T_\Reff}\Big) -1 =& c\m T_\Reff - \frac{a\m}{2} (T\mi-T_\Reff)^2
\alend
where specific heat capacity is relatied to parameter $a\m$ as:
\begeq\label{relationToHeatCapacity}
a\m = \frac{c\m}{T\m}
\eqend
If we assume that room temperature, $T_\Reff$, is at $300$\,K, we observe that the expansion is accurate for a temperature range from $180$\,K to $540$\,K, see Appendix \ref{App1}. We may now introduce above Eq.\,\eqref{tylorExpRHS} into Eq.\,\eqref{defEng} and obtain
\begeq\label{semifinalHelmholtz}
\free\mi = c\m T_\Reff - \frac{a\m}{2} (T\mi-T_\Reff)^2 
+  \frac1{2\rho\mi} \eps\mi_{ij} C\mi_{ijkl} \eps\mi_{kl} 
+ \frac{(T\mi-T_\Reff)}{\rho\mi} \beta\mi_{ij} \eps\mi_{ij} \ .
\eqend

Furthermore, the symmetry of the strain tensor leads to minor symmetries of the stiffness matrix ${C\mi_{ijkl}} = {C\mi_{jikl}} = {C\mi_{ijlk}}$, and without loss of generality the symmetry of thermoelastic interaction ${\beta_{ij}\M} = {\beta_{ji}\M}$, we obtain
\begeq\label{finalHelmholtz}
\free\mi = c\m T_\Reff - \frac{a\m}{2} (T\mi-T_\Reff)^2 
+  \frac1{2\rho\mi} u\mi_{i,j} C\mi_{ijkl} u\mi_{k,l} 
+ \frac{(T\mi-T_\Reff)}{\rho\mi} \beta\mi_{ij} u\mi_{i,j} \ .
\eqend 
The equations are closed such that thermoelastic material is modeled at microscale by means of Eq.\,\eqref{entropy}, as follows:
\begal
\eta\mi =& -\pd{\free\mi}{T\mi} = a\m (T\mi-T_\Reff) + \frac{1}{\rho\mi} \beta\mi_{ij} u\mi_{i,j} \, , \\
\sigma\mi_{ji} =& \rho\mi \pd{\free\mi}{\eps\mi_{ij}} = C\mi_{ijkl} u\mi_{k,l} + (T\mi-T_\Reff) \beta\mi_{ij} \ .
\alend
Furthermore, we consider steady state condition for temperature and displacement by setting their rate terms equal to zero,
\begeq\label{smallStrain}
    \bigg( C\mi_{ijkl} u\mi_{k,l} + \beta\mi_{ij}(T\m-T_\Reff)  \bigg)_{,j} + \rho\mi g_i = 0 \ , \\
    - q\mi_{i,i} - \rho\mi r = 0 \ .
\eqend 
For the homogenized continuum, we employ one axiom that the free energy within the RVE, $\Omega$, is identical at micro- and macroscale as:
\begeq\label{hEngComparison}
\int_\Omega \free\mi \dd V = \int_\Omega \free\ma \dd V \ .
\eqend
Furthermore, we simplify the analysis by assuming a mass ratio for the following terms:
\begeq\label{ratio}
\rho\ma \int_\Omega \dd V =  \int_\Omega \rho\mi \dd V \, , \
V = \int_\Omega \dd V \, , \
\rho\ma  =  \frac1{V} \int_\Omega \rho\mi \dd V \, , \
\eqend

\subsection{Macroscale}
A higher-order macroscale model may be defined by strain $\ten\varepsilon\M$, gradient of strain $\nabla\ten\varepsilon\M$, and temperature $ T\M$. In other words, we begin with a specific free energy, $\free\M (\varepsilon\M_{ij}, \varepsilon\M_{ij,k},  T \M)$, where we use the comma notation as the partial space derivative. We emphasize that the microstructure causes higher order in displacement because of homogenization of the structure \cite{075}; however, we exclude temperature gradient from the free energy. We stress that temperature gradient is used in heat flux as a consequence of the second law of thermodynamics. Free energy is obtained from the internal energy, in our formulation, internal energy incorporates reversible quantities. \\

We use the simplest possible thermo-mechanical model in generalized mechanics. A reference thermo-mechanical state, $\ten\varepsilon\M_\text{ref} = 0$, $\nabla\ten\varepsilon\M_\text{ref} = 0$, and $T\M=T\M_\text{ref}$, is considered and kinematic, balance and constitutive equations are linearized with respect to the reference state. Macroscale Helmholtz free energy, $\free\M (\varepsilon\M_{ij}, \varepsilon\M_{ij,k},  T \M)$, is then specified as a quadratic form,
\begin{align}\label{macroHEng}
\begin{split}
    \rho\M \free\M (\varepsilon\M_{ij}, \varepsilon\M_{ij,k},  T \M) = &
    c\M T_\Reff+\frac{1}{2}C\M_{ijkl} u\M_{i,j} u\M_{k,l} 
    - \beta\M_{ij} u\M_{i,j}( T \M- T_\text{ref})
    \\ &
    +G\M_{ijklm} u\M_{i,j} u\M_{k,lm} 
    + \frac{1}{2}D\M_{ijklmn} u\M_{i,jk} u\M_{l,mn} 
    \\&
    -\frac{1}{2}a\M( T\M -  T_\text{ref})^2
    +\gamma\M_{ijk} u\M_{i,jk}( T\M -  T_\text{ref}) \ ,
    \end{split}
\end{align}
where we have used the symmetry of strain, $\varepsilon\M_{ij}=(u\M_{i,j}+u\M_{j,i})/2$, allowing us to consider additional minor symmetries ${G\M_{ijklm}} = {G\M_{jiklm}} = {G\M_{ijkml}} = {G\M_{lmijk}}$ and ${D\M_{ijklmn}} = {D\M_{jiklmn}} = {D\M_{ijkmln}} = {D\M_{lmnijk}}$, with the usual restrictions for positive definiteness \cite{nazarenko2021uniqueness,nazarenko2021positive}, and without loss of generality the symmetry of ${\gamma_{ijk}\M} = {\gamma_{kji}\M}$. In analogy with Eq.\,\eqref{balancEntropy1}, the governing equations at the macroscale read by following a variational formulation \cite{030}, as follows:
\begeq\label{governingMacro}
\rho\ma (\ddot{u}\ma_i) -  \bigg( \rho\ma \pd{\free\ma}{\eps_{ij}} \bigg)_{,j} + \bigg( \rho\ma \pd{\free\ma}{\eps_{ij,k}} \bigg)_{,jk}  - \rho\ma g_i = 0 \ , \\
\rho\ma \Dt{ \bigg( \pd{\free\ma}{T\ma} \bigg)} - \frac1{T\ma} q\ma_{i,i} + \rho\ma \frac{r}{T\ma} = 0 \ .
\eqend
Now, by using the model in Eq.\,\eqref{macroHEng} for the free energy, in the case of steady state as in Eq.\,\eqref{smallStrain}, we obtain
\begal\label{governingMacro2}
&-  \bigg(  C\M_{ijkl} u\M_{k,l} -\beta\M_{ij} (T\M-T_\text{ref}) + G\M_{ijklm} u\M_{k,lm}  \bigg)_{,j} 
+\\
& \quad + \bigg( G\M_{lmijk} u\M_{l,m}  \bigg)_{,jk} 
- \rho\ma g_i = 0 \ ,
\\
&- q\ma_{i,i} + \rho\ma r = 0 \ .
\alend
The main aim is to find a relation between microscale and macroscale parameters. In other words, we start with given parameters in Eq.\,\eqref{smallStrain} and obtain the parameters in Eq.\,\eqref{governingMacro2}. 

We introduce a geometric center of the RVE, $\overset{c}{\ten{X}}$, as follows:
\begin{equation}\label{geomCen}
    \overset{c}{\ten{X}} = \frac{1}{V}\int_{\Omega}\ten{X} \d V \, , \
    V = \int_{\Omega} \d V \ .
\end{equation}
Assuming displacement and temperature field $\ten{u}\M$ and $ T \M$ are continuous over the microscale, we approximate the macroscale displacement and temperature by a Taylor expansion around the value at the geometric center and truncate terms with orders higher than quadratic for displacement, since we assume that the energy depends up to the second gradient \cite{yang2019determination}, and truncate terms with orders higher than linear for temperature, since the energy depends on temperature, but not on its gradient. Macroscopic displacement field and its displacement gradients read
\begal\label{dispTaylor}
    u_{i}\M(\mathbf{X}) =& u_{i}\M\Big|_{\overset{c}{\ten X}} + u_{i,j}\M\Big|_{\overset{c}{\ten{X}}}({X}_j-\overset{c}{X}_j) + \frac{1}{2}u_{i,jk}\M\Big|_{\overset{c}{\ten X}}({X}_j-\overset{c}{X}_j)({X}_k-\overset{c}{X}_k)\\
    u_{i,l}\M(\mathbf{X}) =& u_{i,l}\M\Big|_{\overset{c}{\ten X}} +  u_{i,lk}\M\Big|_{\overset{c}{\ten X}}({X}_k-\overset{c}{X}_k)\\
    u_{i,lm}\M(\mathbf{X}) =& u_{i,lm}\M\Big|_{\overset{c}{\ten X}} \ ,
\alend
since $(\cdot)|_{\overset{c}{\ten X}}$ is evaluated at the geometric center and thus a constant vanishing by taking its derivative. We stress that there is no scale separation such that the gradient at macroscale is used in this expansion by means of the comma notatopn. Macroscopic temperature field is assumed constant over the RVE (at microscale) leading to 
\begin{equation}\label{tempExp}
    T \M(\mathbf{X}) =  T \M(\overset{c}{\ten X}) \ .
\end{equation}
In Eq.\,\eqref{dispTaylor}, the first and second derivatives of macroscopic deformation field are the unknowns. They are obtained by spatial averaging and exploiting the fact that terms evaluated at $\overset{c}{\ten{X}}$ are constant within $\Omega$, thus,
\begin{align}\label{avrgDisp}
    \begin{split}
    & \langle u_{i,j}\M\rangle = \frac{1}{V}\int_{\Omega}u_{i,j}\M \d V = u_{i,j}\M\Big|_{\overset{c}{X}} +  u_{i,jk}\M\Big|_{\overset{c}{X}} \underbrace{\frac{1}{V}\int_{\Omega}(X_k-\overset{c}{X}_k)dV}_{\overset{c}{\ten{X}}-\overset{c}{\ten{X}}\text{= 0}} = u_{i,j}\M\Big|_{\overset{c}{X}}\\
    &\langle u_{i,jk}\M\rangle = \frac{1}{V}\int_{\Omega}u_{i,jk}\M \d V = u_{i,jk}\M\Big|_{\overset{c}{X}} \ .
    \end{split}
\end{align}
Going back to the Eq.\,\eqref{dispTaylor}, we replace the displacement gradients with spatially averaged values from Eq.\,\eqref{avrgDisp}, as follows:
\begin{align}\label{macroDisp}
    \begin{split}
    &  u_{i,j}\M(\mathbf{X}) = \langle u_{i,j}\M\rangle +  \langle u_{i,jk}\M\rangle({X}_k-\overset{c}{X}_k)\\
    &  u_{i,jk}\M(\mathbf{X}) = \langle u_{i,jk}\M\rangle \ .
    \end{split}
\end{align}
We use the axiom in Eq.\,\eqref{hEngComparison} and insert the above averages into the macroscale energy definition on the right-hand side of Eq.\,\eqref{macroHEng}. All the spatial averaged terms are constant within the RVE such that they are taken out of the integral. For the sake of clarity, we write each term of the free energy at macroscale by denoting the corresponding parameter to be determined
\begal\label{macroParameters}
\ten c\M \Rightarrow&\int_{\Omega} c\M T_\Reff \d V = c\M T_\Reff V\\
\ten C\M \Rightarrow&\int_{\Omega}\frac{1}{2} C_{ijlm}\M \big(\langle u\M_{i,j}\rangle + \langle u\M_{i,jk}\rangle(X_k - \cen{X}_k)\big)\big(\langle u\M_{l,m}\rangle + \langle u\M_{l,mn}\rangle(X_n - \cen{X}_n)\big) \d V = \\
&=\frac{1}{2}C\M_{ijlm}\Big( \langle u\M_{i,j}\rangle\langle u\M_{l,m}\rangle + I_{kn}\langle u\M_{i,jk}\rangle\langle u\M_{l,mn}\rangle\Big) V \\
\ten D\M \Rightarrow&\int_{\Omega}\frac{1}{2} D\M_{ijklmn}\langle u\M_{i,jk}\rangle\langle u\M_{l,mn}\rangle \d V = \frac{1}{2}D\M_{ijklmn} \langle u\M_{i,jk}\rangle\langle u\M_{l,mn}\rangle V \\  
\ten G\M \Rightarrow&\int_{\Omega} G\M_{ijklm}\big(\langle u\M_{i,j}\rangle + \langle u\M_{i,jk}\rangle(X_k - \cen{X}_k)\big)\langle u\M_{k,lm}\rangle \d V =G\M_{ijklm} \langle u\M_{i,j}\rangle\langle u\M_{k,lm}\rangle V \\  
\ten a\M \Rightarrow&\int_{\Omega}\frac{1}{2} a\M \big(T\M - T_\Reff\big)\big(T\M - T_\Reff\big) \d V = \frac{1}{2}a\M\big(T\M - T_\Reff\big)^2V \\
\ten{\beta}\M\Rightarrow&\int_{\Omega} \beta\M_{ij} \big( \langle u\M_{i,j}\rangle + \langle u\M_{i,jk}\rangle(X_k - \cen{X}_k) \big)\big(T\M - T_\Reff \big) \d V =\beta_{ij}\M \langle u_{i,j}\M\rangle\big( T\M - T_\Reff\big)V\\
\ten{\gamma}\M \Rightarrow&\int_{\Omega}\gamma\M_{ijk}\langle u_{i,jk}\M\rangle \big(T\M - T_\Reff\big) \d V = \gamma_{ijk}\M \langle u_{i,jk}\M\rangle (T\M - T_\Reff)V \\
\alend
where
\begal\label{identityMatrx}
I_{kn} = \int_{\Omega} (X_{k} - \cen{X}_{k})(X_{n} - \cen{X}_{n}) \d V \ .
\alend
Separating above equations and combining parameters with identical combinations of spatial averages, we obtain macroscale deformation energy,
\begal\label{finalMacroenergy}
    \int_{\Omega}\free\M \d V = &
        \frac{V}{2}\Bigg(
        2c\M T_\Reff 
        + C_{ijlm}\M \langle u_{i,j}\M\rangle \langle u_{l,m}\M\rangle 
        - 2\beta_{ij}\M\langle u_{i,j}\M\rangle(T \M -  T _\text{ref})\\
        &+ 2 G_{ijklm}\M \langle u_{i,j}\M\rangle \langle u_{k,lm}\M\rangle 
    + \bigg(C_{ijlm}\M I_{kn} + D_{ijklmn}\M \bigg)\langle u_{i,jk}\M\rangle \langle u_{l,mn}\M\rangle\\
    &+2\gamma_{ijk}\M\langle u_{i,jk}\M\rangle(T \M -  T _\text{ref}) - a\M(T \M -  T _\text{ref})^2\Bigg)
\alend
\subsection{Microscale free energy}
\begin{figure}[!h]
    \centering
    \includegraphics[scale = 0.45]{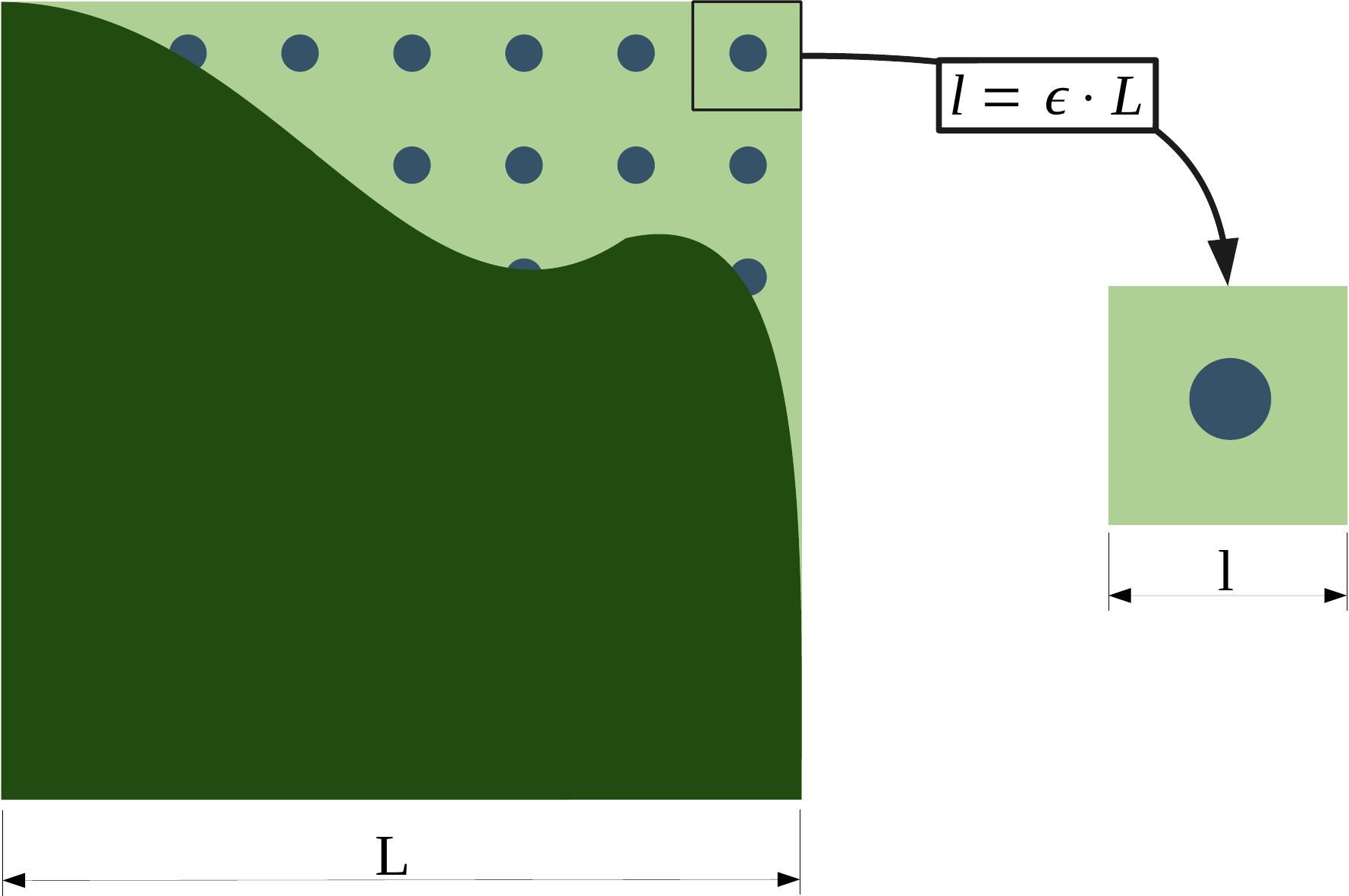}
    \caption{Scaling from macroscale length L to microscale length l via homothetic ration $\epsilon$}
    \label{homotheticRatio}
\end{figure}

The basic idea behind this method is to transfer between microscale, $\ten y$, and macroscale, $\ten X$, by means of the so-called homothetic ratio $\epsilon$. We visualize the meaning of this rather abstract constant in Fig.\,\ref{homotheticRatio} and emphasize that it is a constant within the RVE.
In the computation, there is one single coordinate system and we may set $\epsilon=1$ and model the RVE in real geometric dimensions. In general, the homothetic ratio is the connection between 
\begin{equation}\label{homeoRatio}
    \underbrace{\epsilon = \frac{l}{L} = \frac{\text{microscale length}}{\text{macroscale length}}}_{\text{Homothetic ratio}} 
    \, , \   
    \underbrace{ y_j = \frac{1}{\epsilon}(X_j - \overset{c}{X}_j)}_{\text{Local coordinate}}
    \, ,
\end{equation}
such that we obtain $y_{i,j} = \delta_{ij}/\epsilon$. Microscale displacement field for the RVE is then expanded with regard to $\epsilon$,
\begin{equation}\label{dispField}
    \ten{\ten{u}}\m(\ten{X}) = \underbrace{\overset{0}{\ten{\ten{u}}}(\ten{X},\ten{y}) + \epsilon \overset{1}{\ten{\ten{u}}}(\ten{X},\ten{y}) + \epsilon^2 \overset{2}{\ten{u}}(\ten{X},\ten{y}) + \mathcal{O}(\epsilon^3)}_{\text{Expanded microscale displacement field}}
\end{equation}
where $\overset{n}{\ten{u}}(\ten{X},\ten{y})$ is $\ten{y}$-periodic as an assertion from the unit cell. Furthermore, we assume that temperature is constant over the RVE,
\begin{equation}\label{tempField}
     T\m(\ten{X}, \ten{y}) =  T\M(\ten{X}) \ ,
\end{equation}
in order to ensure the $\ten y$-periodicity of a scalar field. By using the chain rule, we obtain the first derivative of (microscale) displacement field,
\begeq\label{firstDerivDisp}
    u\m _{i,j} = \overset{0}{u}_{i,j} + \frac{1}{\epsilon}\frac{\p \overset{0}{u}_i}{\p y_j} 
    + \epsilon \bigg(  \overset{1}{u}_{i,j} + \frac{1}{\epsilon} \frac{\p \overset{1}{u}_i}{\p y_j} \bigg) 
    + \epsilon^2 \bigg( \overset{2}{u}_{i,j} + \frac{1}{\epsilon}\frac{\p \overset{2}{u}_i}{\p y_j} \bigg) + \mathcal{O}(\epsilon^3) \ .
\eqend
We utilize Eq.\,\eqref{smallStrain}, use chain rule, and insert Eq.\,\eqref{firstDerivDisp},
\begal\label{expandedConstitutive}
    &\Bigg( 
    {C_{ijkl}\m }\bigg( \overset{0}{u}_{k,l} + \frac{1}{\epsilon}\frac{\p \overset{0}{u}_k}{\p y_l} + \epsilon\overset{1}{u}_{k,l} + \frac{\p \overset{1}{u}_k}{\p y_l} + \epsilon^2\overset{2}{u}_{k,l} + \epsilon\frac{\p \overset{2}{u}_k}{\p y_l} \bigg)
    -\beta\m_{ij}\bigg(T\M -  T_\Reff \bigg)\Bigg)_{,j}
    \\
    &+ \frac{1}{\epsilon}\frac{\p}{\p y_j}\Bigg( 
    {C_{ijkl}\m } \bigg(  \overset{0}{u}_{k,l} + \frac{1}{\epsilon}\frac{\p \overset{0}{u}_k}{\p y_l} + \epsilon\overset{1}{u}_{k,l} + \frac{\p \overset{1}{u}_k}{\p y_l} + \epsilon^2\overset{2}{u}_{k,l} + \epsilon\frac{\p \overset{2}{u}_k}{\p y_l}\bigg)
    -\beta\m_{ij}\bigg(T\M -  T_\Reff\bigg) \Bigg) 
    \\
    &+\rho\m  g_i 
    = 0 \ .
\alend
Comparing coefficients in Eq.\,\eqref{expandedConstitutive} of the same order of $\epsilon$ leads to
\begin{itemize}
    \item $\epsilon^{-2}$
    \begeq\label{epsilon2}
        \frac{\p}{\p y_j}\bigg( {C_{ijkl}\m } \frac{\p \overset{0}{u}_k}{\p y_l}\bigg) = 0 \ .
    \eqend
    \item $\epsilon^{-1}$
    \begal\label{epsilon1}
         &\Big({C_{ijkl}\m } \frac{\p \overset{0}{u}_k}{\p y_l}\Big)_{,j} 
         + \frac{\p}{\p y_j}({C_{ijkl}\m } \overset{0}{u}_{k,l}) 
         + \frac{\p}{\p y_j}\Big({C_{ijkl}\m } \frac{\p \overset{1}{u}_k}{\p y_l}\Big)\\
         &- \frac{\p}{\p y_j}\Big(\beta\m_{ij}(T\M -  T_\Reff)\Big)= 0 \ .
    \alend
    \item $\epsilon^0$
    \begal\label{epsilon0}
        &\big({C_{ijkl}\m } \overset{0}{u}_{k,l} \big)_{,j} 
        + \Big({C_{ijkl}\m } \frac{\p \overset{1}{u}_k}{\p y_l}\Big)_{,j} 
        + \frac{\p}{\p y_j}\Big( {C_{ijkl}\m } \overset{1}{u}_{k,l} \Big) 
        + \frac{\p}{\p y_j}\Big( {C_{ijkl}\m } \frac{\p \overset{2}{u}_k}{\p y_l} \Big)\\
        &- \big(\beta\m_{ij}T\M \big)_{,j} + \rho\m  g_i = 0 \ .
    \alend
    \item $\epsilon^1$
    \begeq
    \bigg( C\m_{ijkl} \Big( \overset{1}{u}_{k,l} + \pd{ \overset{2}{u}_k }{y_l} \Big) \bigg)_{,j}
    + \pd{}{y_j} \bigg( C\m_{ijkl} \overset{2}{u}_{k,l} \bigg) = 0 \ .
    \eqend
    \item $\epsilon^2$
    \begeq
    \big( C\m_{ijkl}  \overset{2}{u}_{k,l}  \big)_{,j} = 0 \ .
    \eqend
\end{itemize}
Only possible solution for Eq.\,\eqref{epsilon2} is to define $\overset{0}{\ten u}$ as a function of $\ten{X}$ because $\ten{C}\m$ and $\ten{\beta}\m$ depends on local variable $\ten{y}$. This argumentation leads to the conclusion,
\begin{equation}\label{u0disp}
    \overset{0}{u}_i(\ten{X}) = {u}\M_i(\ten{X}) \, .
\end{equation}
We use a separation of variables or also called Bernoulli ansatz and rewrite,
\begal\label{def.u1u2.terms}
\overset{1}{u}_i(\ten X, \ten y) =& u\M_{a,b}(\ten{X}) \varphi_{abi}(\ten y) - (T\M(\ten X) - T_\Reff) P_i(\ten y) \ , \\
\overset{2}{u}_i(\ten X, \ten y) =& u\M_{a,bc}(\ten{X}) \psi_{abci}(\ten y) \ ,
\alend 
where we introduce unknown tensors $\ten\varphi$, $\ten\psi$, $\ten P$ of one rank higher so that the formulation is general. We emphasize that the temperature is only expanded upto one order less than the displacement. By inserting these into Eq.\,\eqref{epsilon1}, we acquire
\begal
     & \frac{\p}{\p y_j}({C_{ijkl}\m } u\M_{k,l}) 
     + \frac{\p}{\p y_j}\Bigg({C_{ijkl}\m } \pd{}{y_l} \Big( u\M_{a,b} \varphi_{abk} - (T\M - T_\Reff) P_k  \Big)\Bigg)
     \\& \quad- \frac{\p}{\p y_j}\Big(\beta\m_{ij}(T\M -  T_\Reff)\Big)= 0 \ , \\
     & u\M_{a,b} \pd{}{y_j} \Bigg( C_{ijkl}\m \Big( \delta_{ak}\delta_{bl} + \pd{\varphi_{abk}}{y_l} \Big) \Bigg)
     - (T\M - T_\Reff) \pd{}{y_j} \Bigg( C_{ijkl}\m \pd{P_k}{y_l} + \beta\m_{ij} \Bigg) = 0 \ .
\alend
The only possible general solution is to fulfill (solve) independently the following governing equations
\begin{equation}\label{u1PDE}
    \frac{\p}{\p y_j}\Bigg( C\m_{ijkl}\bigg(\frac{\p \varphi_{abk}}{\p y_l} +\delta_{ak}\delta_{bl} \bigg) \Bigg) = 0 \, .
\end{equation}
in order to obtain $\ten\varphi$ and 
\begin{equation}\label{TPDE}
    \frac{\p}{\p y_j}\Bigg( C\m_{ijkl}\frac{\p P_{k}}{\p y_l} +\beta\m_{ij} \Bigg) = 0 \, .
\end{equation}
for acquiring $\ten P$. Repeating the same procedure for Eq.\,\eqref{epsilon0}, we have
\begal
    &
    \Bigg(
    C_{ijkl}\m  u\M_{a,b} \delta_{ak}\delta_{bl} 
    + {C_{ijkl}\m } \pd{}{y_l} \Big( u\M_{a,b} \varphi_{abk} - (T\M - T_\Reff) P_k  \Big) \Bigg)_{,j} 
    \\ &
    + \pd{}{y_j} \Bigg( C_{ijkl}\m  \overset{1}{u}_{k,l}  
    + C_{ijkl}\m \pd{}{y_l} \Big( u\M_{a,bc} \psi_{abck} \Big) \Bigg)
    - \big(\beta\m_{ij}T\M \big)_{,j} + \rho\m  g_i = 0 \ ,
\alend
which is rewritten
\begal
    &
    \Bigg(
    u\M_{a,b} C_{ijkl}\m  \Big( \delta_{ak}\delta_{bl} 
    +  \pd{\varphi_{abk}}{y_l} \Big)  - C_{ijkl}\m (T\M - T_\Reff) \pd{P_k}{y_l}  - \beta\m_{ij}T\M \Bigg)_{,j} 
    +\\ &
    + \pd{}{y_j} \Bigg( C_{ijkl}\m  \bigg( 
    \Big( u\M_{a,b} \varphi_{abk} - (T\M - T_\Reff )P_k  \Big)_{,l}  
    +  \pd{}{y_l} \Big( u\M_{a,bc} \psi_{abck} \Big) 
    \bigg) \Bigg)
    + \rho\m  g_i = 0 \ , 
    \\
    &
    u\M_{a,bj} C_{ijkl}\m  \Big( \delta_{ak}\delta_{bl} 
    +  \pd{\varphi_{abk}}{y_l} \Big)  - C_{ijkl}\m T\M_{,j} \pd{P_k}{y_l}  - \beta\m_{ij} T\M_{,j} 
    +\\ &
    + \pd{}{y_j} \Bigg( C_{ijkl}\m  \bigg( 
     u\M_{a,bl} \varphi_{abk} - T\M_{,l} P_k   
    +  u\M_{a,bc} \pd{\psi_{abck}}{y_l}  
    \bigg) \Bigg)
    + \rho\m  g_i = 0 \ , 
    \\
    &
    u\M_{a,bc} \Bigg( 
    \delta_{cj} C_{ijkl}\m  \Big( \delta_{ak}\delta_{bl}  +  \pd{\varphi_{abk}}{y_l} \Big)
    + \pd{}{y_j} \bigg( C_{ijkl}\m  \Big( \delta_{lc} \varphi_{abk} +  \pd{\psi_{abck}}{y_l} \Big) \bigg)
    \Bigg)
    - \\
    &
    -T\M_{,a} \Bigg( C_{iakl}\m \pd{P_k}{y_l} + \beta\m_{ia} + \pd{}{y_j} \bigg( C_{ijkl}\m \delta_{al} P_k \bigg) \Bigg)
    + \rho\m  g_i = 0 \ , 
\alend
Furthermore, from Eq.\,\eqref{governingMacro2}, by inserting 
\begeq
\rho\ma g_i = -  C\M_{ijkl} u\M_{k,lj} +\beta\M_{ij} T\M_{,j} - G\M_{ijklm} u\M_{k,lmj}  
+ G\M_{lmijk} u\M_{l,mjk}  
\eqend
and using the same cut-off procedure for displacement third derivative and temperature second derivative, we obtain
\begal
    & u\M_{a,bc} \Bigg( 
    \delta_{cj} C_{ijkl}\m  \Big( \delta_{ak}\delta_{bl}  +  \pd{\varphi_{abk}}{y_l} \Big)
    + \pd{}{y_j} \bigg( C_{ijkl}\m  \Big( \delta_{lc} \varphi_{abk} +  \pd{\psi_{abck}}{y_l} \Big) \bigg)
    \Bigg)
    - \\
    &
    -T\M_{,a} \Bigg( C_{iakl}\m \pd{P_k}{y_l} + \beta\m_{ia} + \pd{}{y_j} \bigg( C_{ijka}\m P_k \bigg) \Bigg)
    + \frac{\rho\m}{\rho\M}  \Big( -  C\M_{ijkl} u\M_{k,lj} +\beta\M_{ij} T\M_{,j} \Big) = 0 \ , 
\alend    
We separate the independent parts and enforce to fulfill
\begin{equation}\label{u2PDE}
    \frac{\p}{\p y_j}\Bigg( C\m_{ijkl}\bigg(\frac{\p \psi_{abck}}{\p y_l} +\varphi_{abk}\delta_{cl} \bigg) \Bigg) + C\m_{ickl}\bigg(\frac{\p \varphi_{abk}}{\p y_l} +\delta_{ak}\delta_{lb} \bigg) - \frac{\rho\m}{\rho\M}C\M_{icab} = 0 
    \ ,
\end{equation}
in the case of $T\M_{,a}=0$ and
\begeq
C_{iakl}\m \pd{P_k}{y_l} + \beta\m_{ia} + \pd{}{y_j} \bigg( C_{ijka}\m P_k \bigg) - \frac{\rho\m}{\rho\M}\beta\M_{ia} = 0 \ ,
\eqend
in the case of $u\M_{a,bc}=0$. At the moment that we have assumed temperature without higher order contributions, the latter is identically fulfilled. Therefore, we solve Eqs.\,\eqref{u1PDE}, \eqref{TPDE}, \eqref{u2PDE} in order to determine $\ten\varphi$, $\ten P$, $\ten\psi$, respectively, under the condition $T\M_{,a}=0$. 
By introducing these shorthand notations
\begal\label{parametersuGradMicroscaleAVG2}
  L_{abij} = & \delta_{ia}\delta_{jb} + \pd{ \varphi_{abi}}{ y_j} \ ,
  \\
  N_{abcij} = & \varphi_{abi}\delta_{jc} + \pd{ \psi_{abci}}{ y_j} \ ,
  \\
  Z_{ij} = & \frac{\p P_{i}}{\p y_{j}} \ ,
\alend
We need to fulfill
\begal
\varphi_{abi} \Leftarrow &
\pd{}{ y_j}\Bigg( C\m_{ijkl} L_{abkl} \Bigg) = 0  \, , \\
P_i \Leftarrow &
\frac{\p}{\p y_j}\Bigg( C\m_{ijkl} Z_{kl} +\beta\m_{ij} \Bigg) = 0 \, , \\
\psi_{abci} \Leftarrow &
\pd{}{y_j}\Bigg( C\m_{ijkl} N_{abckl} \Bigg) 
+ C\m_{ickl} L_{abkl}
- \frac{\rho\m}{\rho\M}C\M_{icab} = 0 \ , \text{ where } T\M_{,a}=0 \ , 
\alend
By using Eq.\,\eqref{def.u1u2.terms} in Eq.\,\eqref{dispField}, we obtain
\begal\label{uMicroscale}
         u\m_{i}(\ten{X}, \ten{y}) = &
         u\M_{i}(\ten{X}) + \epsilon \Big( \varphi_{abi}(\ten{y})u\M_{a,b}(\ten{X}) - P_{i}(\ten{y})( T \M(\ten{X})- T_\Reff) \Big) 
         \\ 
         & + \epsilon^2 \Big( \psi_{abci}(\ten{y})u\M_{a,bc}(\ten{X}) \Big) \, .   
         \\
        u\m_{i,j}(\ten{X}, \ten{y}) = &
        u\M_{i,j} + \epsilon \Big( \pd{ \varphi_{abi}}{y_j} \frac{1}{\epsilon} u\M_{a,b} + \varphi_{abi} u\M_{a,bj} 
        - \pd{P_{i}}{y_j} \frac{1}{\epsilon} ( T \M- T_\Reff)  - P_i T\M_{,j} \Big)
        \\ 
         & + \epsilon^2 \Big( \pd{\psi_{abci}}{y_j} \frac{1}{\epsilon} u\M_{a,bc} + \psi_{abci} u\M_{a,bcj}  \Big) \, .   
\alend
We aim for making free energies equivalent, 
\begeq
\int_{\Omega} \free\m \d V = \int_{\Omega} \free\M \d V \ ,
\eqend
where the macroscale energy is given in Eq.\,\eqref{finalMacroenergy}, now we want to find an expression for the microscale energy in order to determine the parameters, where $T\M_{,a}=0$. Moreover, third derivative in displacement vanishes as before and we obtain
\begeq
u\m_{i,j} = \bigg(\delta_{ia}\delta_{jb} + \frac{\p \varphi_{abi}}{\p y_j} \bigg)u\M_{a,b} - \frac{\p P_{i}}{\p y_{j}}( T \M -  T_\Reff) + 
    \epsilon u\M_{a,bc}\bigg( \varphi_{abi}\delta_{jc} + \frac{\psi_{abci}}{\p y_j}\bigg) \, ,
\eqend
after inserting Eq.\,\eqref{macroDisp}, we acquire
\begin{align}\label{uGradMicroscaleAVG}
    \begin{split}
        u\m_{i,j} = &\bigg(\delta_{ia}\delta_{jb} + \frac{\p \varphi_{abi}}{\p y_j} \bigg)(\langle u_{a,b}\M\rangle +  \epsilon y_{c}\langle u_{a,bc}\rangle) - \frac{\p P_{i}}{\p y_{j}}(T \M -  T_\Reff) + \\
        &\epsilon \bigg( \varphi_{abi}\delta_{jc} + \frac{\psi_{abci}}{\p y_j}\bigg)\langle u\M_{a,bc}\rangle \, .
    \end{split}
\end{align}
By introducing
\begeq
  M_{abcij} =  y_c L_{abij} + N_{abcij} \ ,
\eqend
Eq.\,\eqref{uGradMicroscaleAVG} is rewritten by using Eq.\,\eqref{parametersuGradMicroscaleAVG2}, as follows:
\begin{equation}\label{uGradMicroscaleAVG2}
    u\m_{i,j} = L_{abij}\langle u_{a,b}\M\rangle + \epsilon M_{abcij}\langle u_{a,bc}\M\rangle - Z_{ij}(T \M -  T_\Reff)
\end{equation}
Using the above equation microscale energy becomes
\begin{align}\label{finalMicroscaleEnergy}
    \begin{split}
        &\int_{\Omega} \free\m dV = \int_{\Omega} \bigg(\frac{1}{2}C_{ijkl}\m u_{i,j}\m u_{k,l}\m + \beta_{ij} u_{i,j}\m( T \M -  T_\Reff)-\frac{1}{2}a\m  ( T \M -  T_\Reff)^2 + c\m T_\Reff \bigg) dV = \\
        &\frac{1}{2}\int_{\Omega}\bigg\{C_{ijkl}\m L_{abij} L_{cdkl} \langle u_{a,b}\M\rangle\langle u_{c,d}\M\rangle + 2\epsilon C_{ijkl}\m L_{abij}M_{cdekl}\langle u_{a,b}\M\rangle\langle u_{c,de}\M\rangle \\ 
        &- 2\Big[C_{ijkl}\m L_{abij}Z_{kl} - \beta_{ij}\m L_{abij}\Big]\langle u_{a,b}\M\rangle(T \M -  T_\Reff) \\
        &+\epsilon^2C_{ijkl}\m M_{abcij}M_{defkl} \langle u_{a,bc}\M\rangle\langle u_{d,ef}\M\rangle\\
        &-2\epsilon\Big[C_{ijkl}\m M_{abcij}Z_{kl} - \beta_{ij}\m M_{abcij}\Big]\langle u_{a,bc}\M\rangle(T \M -  T_\Reff) \\
        &+\Big[C_{ijkl}\m Z_{ij}Z_{kl} - 2\beta_{ij}\m Z_{ij} -a\m \Big](T \M - T_\Reff)^2 + 2c\m T_\Reff \bigg\}dV\\
        &=\frac{V}{2}\bigg(\bar{C}_{abcd}\langle u_{a,b}\M\rangle\langle u_{c,d}\M\rangle +\bar{G}_{abcde}\langle u_{a,b}\M\rangle\langle u_{c,de}\M\rangle - \bar{\beta}_{ab}\langle u_{a,b}\M\rangle(T \M- T_\Reff)\\
        &+ \bar{D}_{abcdef}\langle u_{a,bc}\M\rangle\langle u_{d,ef}\M\rangle - \bar{\gamma}_{abc}\langle u_{a,bc}\M\rangle(\langle T \M\rangle- T_\Reff) + \bar{a}(T \M- T_\Reff)^2 +\bar{c}T_\Reff\bigg)
    \end{split}
\end{align}
where
\begal\label{finalMicroParameters}
        &\bar{C}_{abcd}=\frac{1}{V}\int_{\Omega}C_{ijkl}\m L_{abij} L_{cdkl} \d V\\
        &\bar{G}_{abcde}=\frac{2\epsilon}{V}\int_{\Omega}C_{ijkl}\m L_{abij}M_{cdekl} \d V\\
        &\bar{\beta}_{ab}=\frac{2}{V}\int_{\Omega}\Big[C_{ijkl}\m L_{abij}Z_{kl} - \beta_{ij}\m L_{abij}\Big] \d V\\
        &\bar{D}_{abcdef}=\frac{\epsilon^2}{V}\int_{\Omega}C_{ijkl}\m M_{abcij}M_{defkl} \d V\\
        &\bar{\gamma}_{abc}=\frac{2\epsilon}{V}\int_{\Omega}\Big[C_{ijkl}\m M_{abcij}Z_{kl} - \beta_{ij}\m M_{abcij}\Big] \d V\\
        &\bar{a}=\frac{1}{V}\int_{\Omega}\Big[C_{ijkl}\m Z_{ij}Z_{kl} - 2\beta_{ij}\m Z_{ij}- a\m\Big] \d V\\
        &\bar{c} = \frac{2}{V}\int_{\Omega}c\m \d V
\alend
Comparing microscale energy in Eq.\,\ref{finalMicroscaleEnergy} to macroscale energy in Eq.\,\ref{finalMacroenergy}, we obtain homogenized values
\begal\label{homogenizedValuesMicroMacro}
        &C_{ijkl}\M = \bar{C}_{ijkl}\\
        &G_{ijklm}\M = \frac{\bar{G}_{ijklm}}{2}\\
        &\beta_{ij}\M = \frac{\bar{\beta}_{ij}}{2}\\
        &D_{ijklmn}\M = \bar{D}_{ijklmn} - C_{ijlm}\M I_{kn}\\
        &\gamma_{ijk}\M = -\frac{\bar{\gamma}_{ijk}}{2}\\
        &a\M = - \bar{a}\\
        &c\M = \frac{\bar{c}}{2}
\alend

\subsubsection{Heat conduction}
For the sake of completeness, we additionally calculate homogenized value of thermal conductivity, $\ten\kappa$. We begin with Eq.\,\eqref{governingMacro2} and use Fourier's law, $q_i = \kappa_{ij} T_{,j}$ leading to 
\begeq\label{heatEq}
\Big( \kappa_{ij}\m T _{,j}\m\Big)_{,i} - \rho\m r = 0
\eqend
We follow the same procedure and expand the microscale temperature field for the RVE with the same accuracy up to first order in $\epsilon$ as follows:
\begin{equation}\label{tempExpansion}
     T \m(\ten{X}) = \underbrace{\overset{0}{ T }(\ten{X},\ten{y}) + \epsilon \overset{1}{ T }(\ten{X},\ten{y}) + \mathcal{O}(\epsilon^2)}_{\text{Expanded microscale temperature field}}
\end{equation}
As we are only interested in expansion up to $\overset{1}{ T }$, thermal conductivity is in relation up to the first-order. In other words, we start with Fourier's equation at the microscale and result in Fourier's equation at the macroscale,
\begeq
\Big( \kappa_{ij}\M T_{,j}\M\Big)_{,i} - \rho\M r = 0 \ .
\eqend
Hence, we obtain
\begeq\label{diff.r}
r = \frac1{\rho\M} \Big( \kappa_{ij}\M T_{,j}\M\Big)_{,i}
\eqend
 Substituting local coordinate $\ten{y}$ into Eq.\,\eqref{tempExpansion}, and using the chain rule, we obtain the first derivative of microscale temperature field,
\begeq\label{expandtempDerivative}
T \m _{,j} = \overset{0}{ T }_{,j} + \frac{1}{\epsilon}\frac{\p \overset{0}{ T }}{\p y_j} + \epsilon\overset{1}{ T }_{,j} + \frac{\p \overset{1}{ T }}{\p y_j} + \mathcal{O}(\epsilon^2) \ .
\eqend
Inserting Eq.\,\eqref{expandtempDerivative} into Eq.\,\eqref{heatEq} and again using chain rule, we obtain an asymptotically expanded governing equation:
\begin{equation}\label{expandedGovHeatEq}
\begin{split}
    &\bigg[\kappa\m_{ij}\bigg(\overset{0}{ T }_{,j} + \frac{1}{\epsilon}\frac{\p \overset{0}{ T }}{\p y_j} + \epsilon\overset{1}{ T }_{,j} + \frac{\p \overset{1}{ T }}{\p y_j} + \epsilon^2\overset{2}{ T }_{,j} + \epsilon\frac{\p \overset{2}{ T }}{\p y_j}\bigg)\bigg]_{,i}\\
    &+ \frac{1}{\epsilon}\frac{\p}{\p y_i}\bigg[\kappa\m_{ij}\bigg(\overset{0}{ T }_{,j} + \frac{1}{\epsilon}\frac{\p \overset{0}{ T }}{\p y_j} + \epsilon\overset{1}{ T }_{,j} + \frac{\p \overset{1}{ T }}{\p y_j} + \epsilon^2\overset{2}{ T }_{,j} + \epsilon\frac{\p \overset{2}{ T }}{\p y_j}\bigg)\bigg] -\rho\m r = 0
\end{split}
\end{equation}
Comparing coefficients in Eq.\,\eqref{expandedGovHeatEq} of the same order of $\epsilon$ leads to
\begin{itemize}
    \item $\epsilon^{-2}$
    \begin{equation}\label{epsilonT2}
        \frac{\p}{\p y_i}\bigg( {\kappa_{ij}\m } \frac{\p \overset{0}{ T }}{\p y_j}\bigg) = 0 \, ,
    \end{equation}
    \item $\epsilon^{-1}$
    \begin{equation}\label{epsilonT1}
         \bigg({\kappa_{ij}\m } \frac{\p \overset{0}{ T }}{\p y_j}\bigg)_{,i} 
         + \frac{\p}{\p y_i} \Big( {\kappa_{ij}\m } \overset{0}{ T }_{,j} \Big) 
         + \frac{\p}{\p y_i}\bigg( {\kappa_{ij}\m } \frac{\p \overset{1}{ T }}{\p y_j}\bigg) = 0\, ,
    \end{equation}
    \item $\epsilon^0$
    \begin{equation}\label{epsilonT0}
        \big({\kappa_{ij}\m } \overset{0}{ T }_{,j} \big)_{,i} 
        + \bigg( {\kappa_{ij}\m } \frac{\p \overset{1}{ T }}{\p y_j}\bigg)_{,i} 
        + \frac{\p}{\p y_i}\big( {\kappa_{ij}\m } \overset{1}{ T }_{,j} \big) 
        - \rho\m  r = 0 \, ,
    \end{equation}
\end{itemize}
Only possible solution for Eq.\,\eqref{epsilonT2} is to define $\overset{0}{ T }$ as a function of $\ten{X}$ because $\kappa\m_{ij}$ depends on local variable $\ten{y}$. This observation leads to a straightforward conclusion,
\begin{equation}\label{T0Eq}
    \overset{0}{ T }(\ten{X}) = { T }\M(\ten{X}) \, , \, 
    \overset{1}{T}(\ten X, \ten y) = -R_j(\ten y) T\M_{,j}(\ten X) \ .
\end{equation}
By using Eq.\,\eqref{T0Eq} in Eq.\,\eqref{epsilonT1}, we obtain
\begeq \label{T1PDE}
\pd{}{y_i} \bigg( \kappa\m_{ij}  T\M_{,j} - \kappa\m_{ij} \pd{ R_k }{y_j} T\M_{,k} \bigg) = 0 \ ,
\\
\pd{}{y_i} \bigg( \kappa\m_{ij}  - \kappa\m_{ik} \pd{ R_j }{y_k}  \bigg) T\M_{,j} = 0 \ ,
\\
\pd{}{y_i} \bigg( \kappa\m_{ij}  - \kappa\m_{ik} \pd{ R_j }{y_k}  \bigg) = 0 \ ,
\eqend 
since $T\M=T\M(\ten X)$. This governing equation reads $R_i$ as the solution. Inserting Eq.\,\eqref{T0Eq} into Eq.\,\eqref{epsilonT0}, we get the following expression:
\begeq\label{thermalPDESolution0}
\Big( \kappa_{ij}\m   T\M_{,j} - \kappa_{ip}\m  \pd{ R_j}{y_p} T\M_{,j} \Big)_{,i} 
- \pd{}{y_i} \Big( \kappa_{ij}\m  T\M_{,aj} R_a \Big) 
- \rho\m  r = 0 \, ,
\\
\Big( \kappa_{ij}\m   T\M_{,j} - \kappa_{ip}\m  \pd{ R_j}{y_p} T\M_{,j} \Big)_{,i} 
- \pd{}{y_i} \Big( \kappa_{ij}\m  T\M_{,aj} R_a \Big) 
- \frac{\rho\m}{\rho\M}  \Big( \kappa_{ij}\M T\M_{,j} \Big)_{,i} = 0 \, ,
\eqend
where we have utilized Eq.\,\eqref{diff.r}.  Since the second gradient in $T\M$ has been vanishing, we obtain
\begeq\label{KapaHomogenized}
\kappa_{ij}\M
= \rho\M  \int_\Omega \frac{1}{\rho\m} \bigg( \kappa_{ij}\m   - \kappa_{ip}\m  \pd{ R_j}{y_p} \bigg) \d V \ .
\eqend

\subsection{Numerical implementation in FEniCS}
Calculation of macroscale parameters in Eq.\,\eqref{homogenizedValuesMicroMacro} and Eq.\,\eqref{KapaHomogenized} requires the solution of $P_i$, $\varphi_{abi}$, $\psi_{abci}$, and $R_j$ tensors from Eq.\,\eqref{TPDE}, Eq.\,\eqref{u2PDE}, Eq.\,\eqref{u1PDE}, and Eq.\,\eqref{T1PDE}.  The computational work has two steps. First, an RVE, $\Omega$, is created in Salome, which has periodic boundary conditions. The solutions of $y$-periodic fields, $P_i$, $\varphi_{abi}$, $\psi_{abci}$, and $R_j$, have to be periodic. Corresponding surfaces must have the identical mesh such that the solution is restricted by this periodicity. In order to attain this condition adequately, projection method is used in Salome for ensuring that the node positions on corresponding boundaries are matching. Second, the weak form is implemented in a Python code to be solved by the open-source packages developed by the FEniCS project. The weak form is obtained by the standard variational formulation by multiplying the governing equation by an arbitrary test function and integrating by parts in order to reduce the regularity condition of the discrete functions. Discretization for the finite element method (FEM) is established by polynomial form functions. We use the same form functions for the fields and their test functions as known as the Galerkin procedure. For representing a vector, for example, $P_i$ in 2D $i=1,2$, we use the Hilbertian Sobolev space, $\mathscr{H}^n$, is of polynomial order, $n$,
\begeq
\mathscr{V} =\Bigg\{ \big\{ P_1, P_2 \big\} \in [ \mathscr{H}^{n}(\Omega) ]^\text{DOF} : \big\{ P_1, P_2 \big\} 
 = \text{given} \ \forall \ten x \in \partial\Omega_\text{D} \Bigg\} \ .
\eqend
hence, we use standard (continuous) Lagrange elements in the FEM \cite{zohdi2018finite}. As known as the Galerkin approach, we use the same type of a functional space for test functions,
\begeq
\bar{\mathscr{V}} =\Bigg\{ \big\{ \del w_1, \del w_2 \big\} \in [ \mathscr{H}^{n}(\Omega) ]^\text{DOF} : \big\{ \del w_1, \del w_2 \big\} 
 = 0 \ \forall \ten x \in \partial\Omega_\text{D} \Bigg\} \ ,
\eqend
where we skip testing the solution at Dirichlet boundaries, $\Omega_\text{D}$, with the known solution. The computational domain, $\Omega$, is the image of the RVE with the Dirichlet type boundary conditions, $\Omega_\text{D}$, being basically periodic boundaries for all fields. \\

\begin{figure}[h!]
    \centering
    \includegraphics[scale = 0.4]{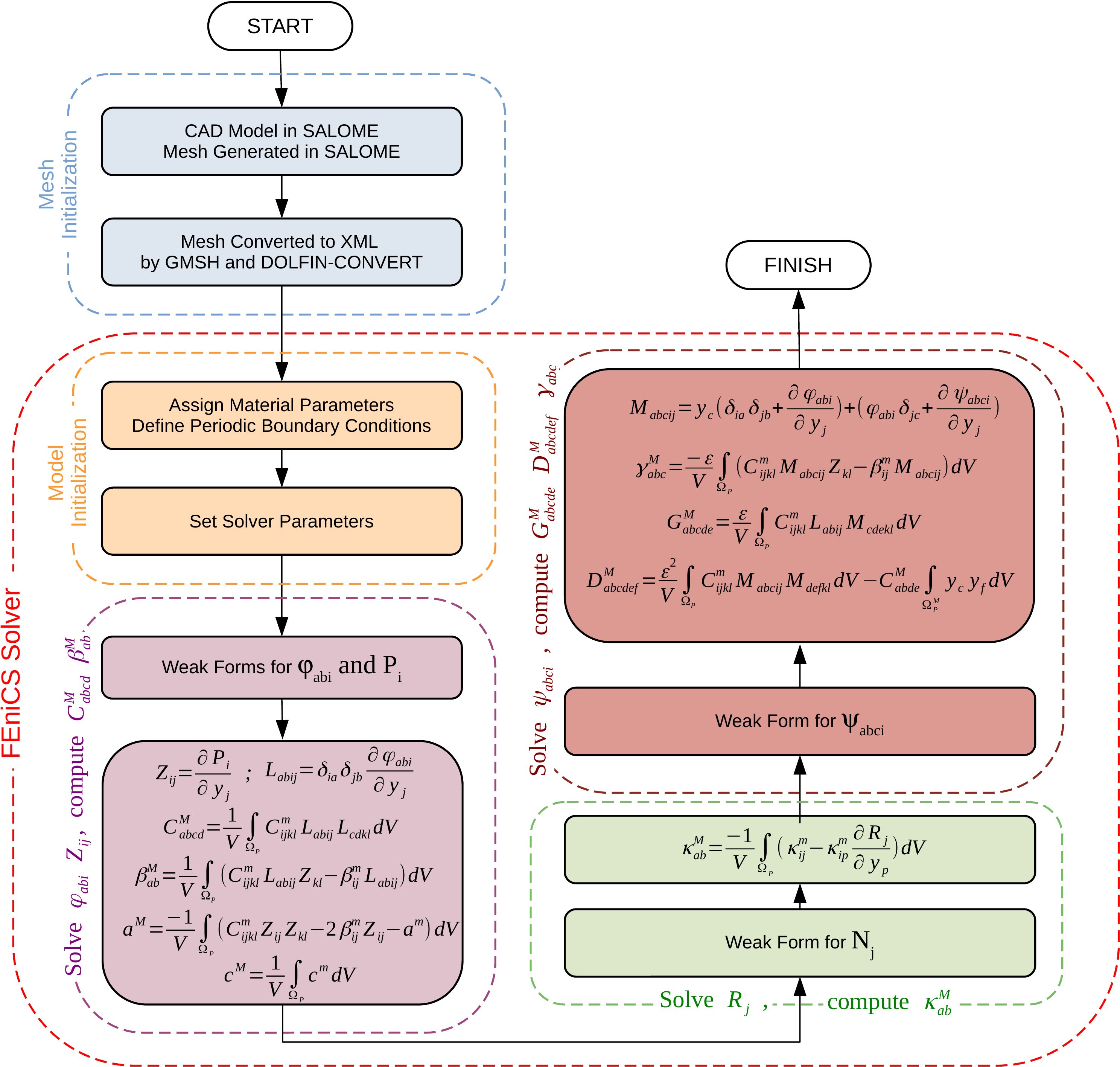}
    \caption{Workflow describing model initialization and solution of the model in FEniCS Solver}
    \label{fenicsWorkflow}
\end{figure}

This calculation is done by solving the corresponding weak forms for governing equations in Eq.\,\eqref{TPDE}, Eq.\,\eqref{u2PDE}, Eq.\,\eqref{u1PDE}, and Eq.\,\eqref{T1PDE}, respectively,
\begal
P_i \Leftarrow & 
\int_\Omega \Bigg( C\m_{ijkl}\frac{\p P_{k}}{\p y_l} +\beta\m_{ij} \Bigg) \del w_{i,j} \d V = 0 \ , 
\\
\psi_{abci} \Leftarrow & 
\int_\Omega \Bigg( 
-\Big( C\m_{ijkl} \frac{\p \psi_{abck}}{\p y_l} + C\m_{ijkl} \varphi_{abk}\delta_{cl} \bigg)   \del w_{i,j} 
+ C\m_{ickl}\bigg(\frac{\p \varphi_{abk}}{\p y_l} +\delta_{ak}\delta_{lb} \bigg) \del w_i -
\\
& \quad - \frac{\rho\m}{\rho\M}C\M_{icab} \del w_i \Bigg) \d V = 0 
    \ , 
\\
\varphi_{abi} \Leftarrow & 
\int_\Omega C\m_{ijkl}\bigg(\frac{\p \varphi_{abk}}{\p y_l} +\delta_{ak}\delta_{bl} \bigg) \del w_{i,j} \d V = 0 \ , 
 \\
R_j \Leftarrow & 
\int_\Omega \bigg( \kappa\m_{ij}  - \kappa\m_{ik} \pd{ R_j }{y_k}  \bigg) \del w_{j,i}= 0 \ .
\alend
Solutions of these fields are then used to construct the macroscale parameters as shown in Fig.\,\ref{fenicsWorkflow}.\\ 

\section{Problem Description}
\begin{figure}
     \centering
     \begin{subfigure}{.45\textwidth}
         \centering
         \includegraphics[width=.8\linewidth]{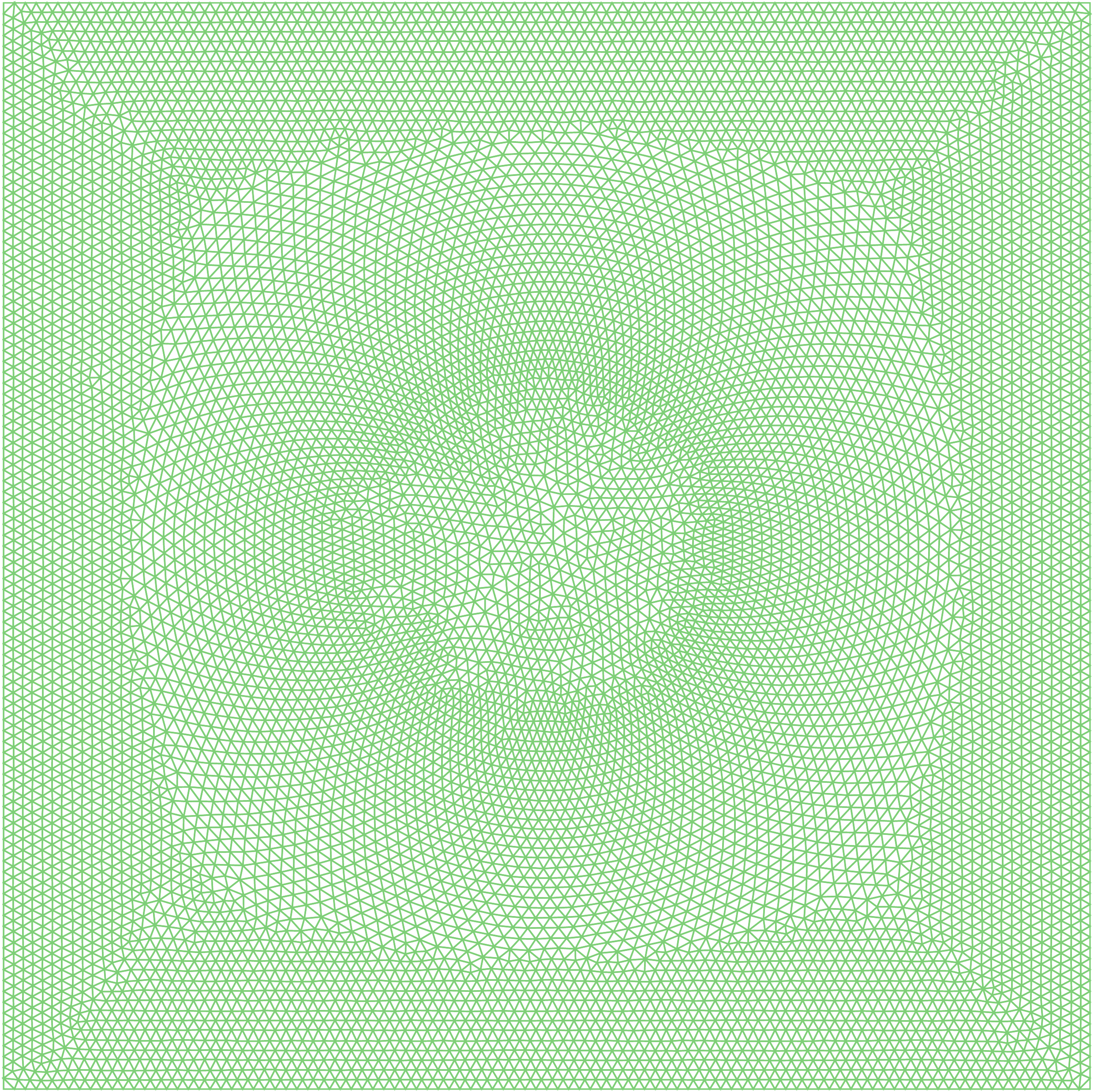}
         \caption{Homogeneous RVE \vspace{0.2in}}
         \label{homog}
     \end{subfigure}\hfill
     \begin{subfigure}{.45\textwidth}
         \centering
         \includegraphics[width=.8\linewidth]{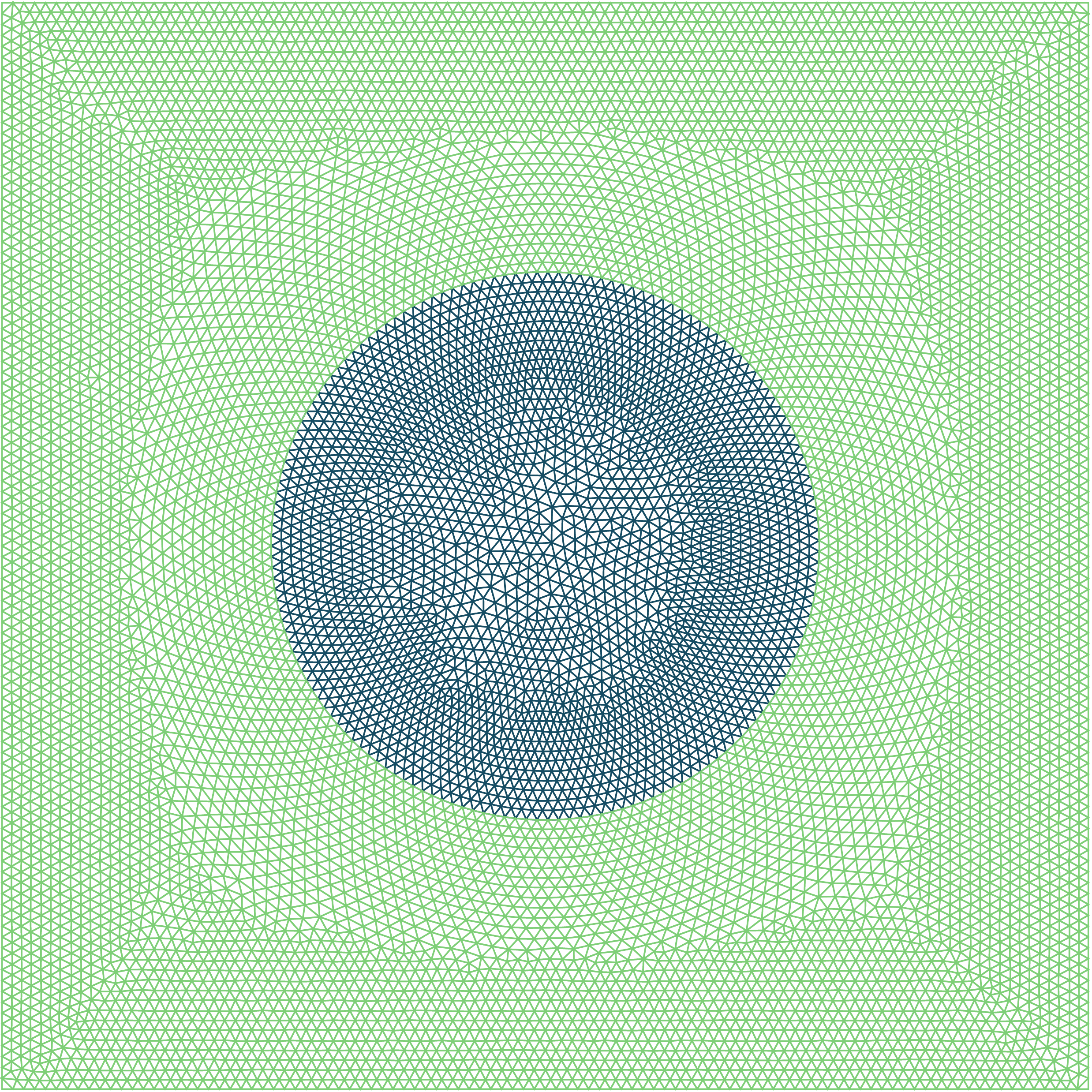}
         \caption{Single central circular pore RVE \vspace{0.2in}}
         \label{single}
     \end{subfigure}
     
     \begin{subfigure}{.45\textwidth}
         \centering
         \includegraphics[width=.8\linewidth]{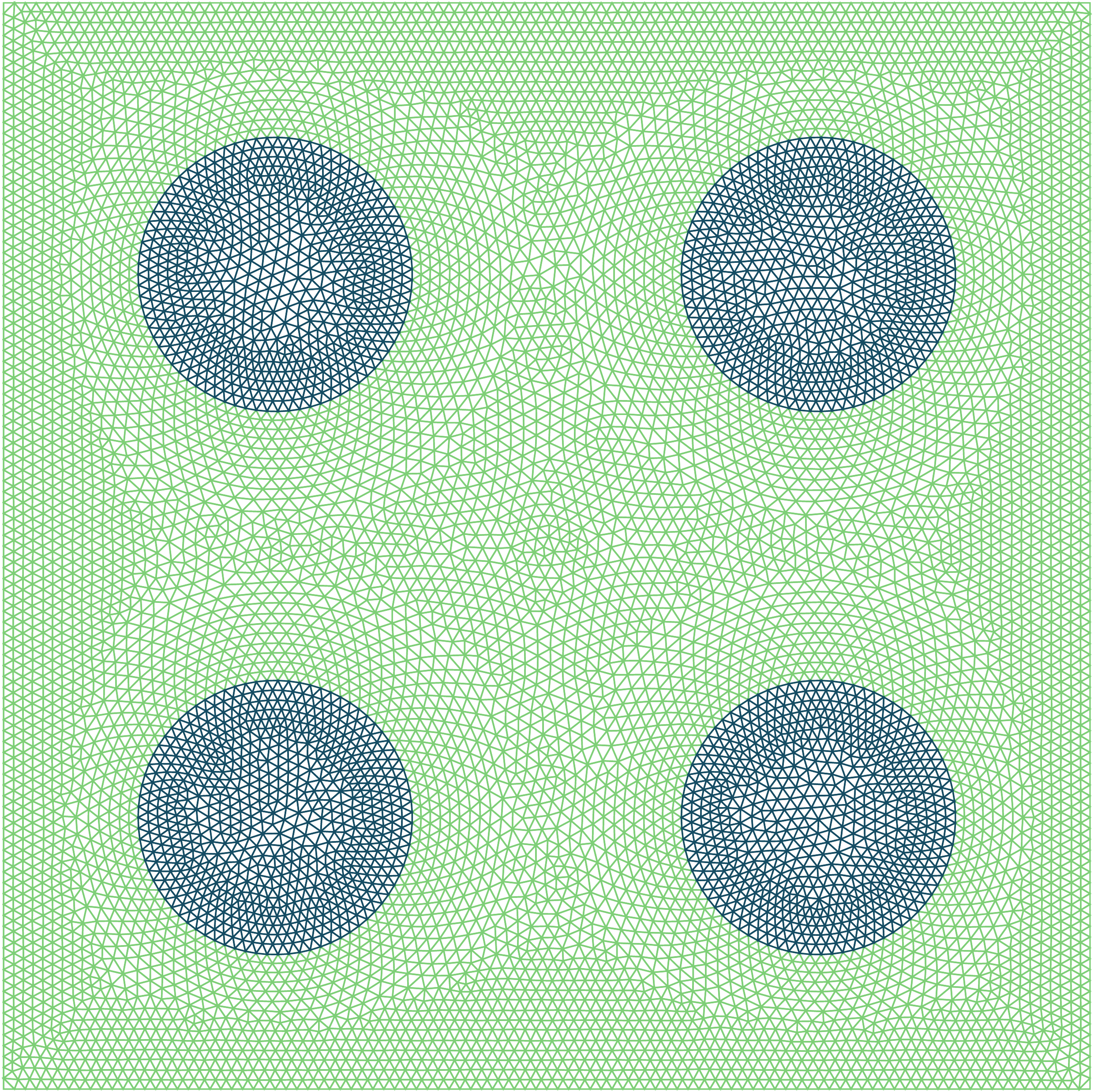}
         \caption{Four uniformly distributed circular pores RVE}
         \label{uniform}
     \end{subfigure}\hfill
     \begin{subfigure}{.45\textwidth}
         \centering
         \includegraphics[width=.8\linewidth]{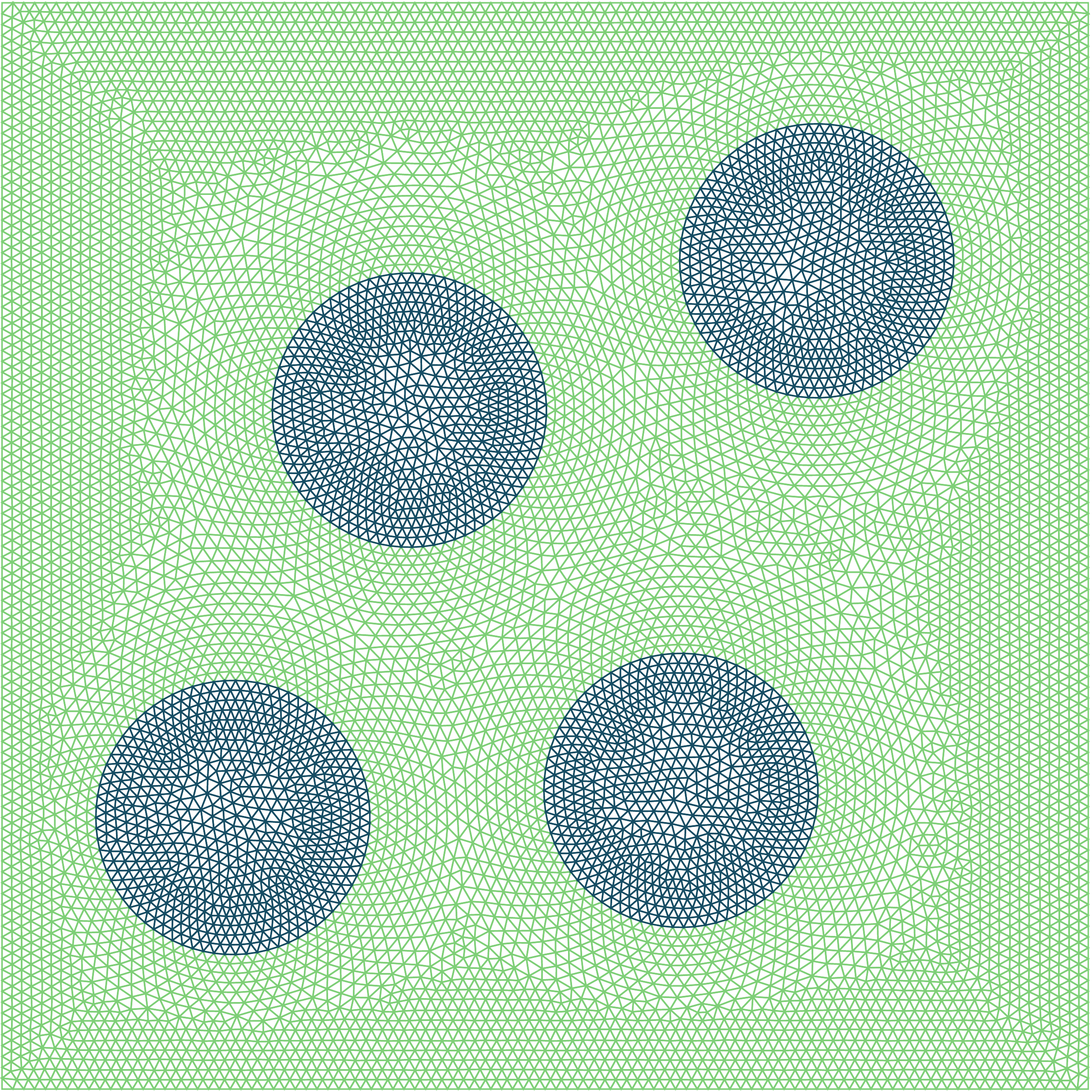}
         \caption{Four randomly distributed circular pores RVE}
         \label{random}
     \end{subfigure}
        \caption{RVEs used for testing of the methodology. Case a) is a homogeneous RVE and cases b), c), and d) are porous cases with different circluar pore distributions at 20\% porosity}
        \label{Solvedexamples}
\end{figure}

Two types of problems have been designed to study the behavior of homogenized material parameters:
\begin{itemize}
\item 
The first problem is intended as a homogeneous material in order to verify that higher-order parameters vanish, $\mathbb{G}\M=0$, $\mathbb{D}\M=0$, and $\mathbb{\gamma}\M=0$, see Figure \ref{homog}. 
\item
The second problem is defined as a porous material with three types of circular pore distributions, Figure \ref{Solvedexamples}. All the structures have the same porosity of 20\%. The idea behind the second problem has been to use different pore distributions to capture previously examined behavior of the higher-order parameters, \cite{vazic2021mechanical}. 
\end{itemize}
The material used in all cases is aluminum, for which we assume a linear elastic material model at the microscale, with material properties compiled in Table \ref{Tab1}. For parameter $a\M$ which is related to heat capacity as in Eq.\,\eqref{relationToHeatCapacity}, we define for all of the cases $T\M = 400$\,K.

\begin{table}[h!]
    \centering
       \renewcommand{\arraystretch}{1.5}
    \begin{tabular}{| c | c |}
    \hline
         Young's modulus, $E$ & $75.0$\,GPa\\
         \hline
         Poisson's ratio, $\nu$ & $0.33$\\
         \hline
         Mass density, $\rho$ & $2700.0$\,m$^3$/kg\\
         \hline
         Thermal expansion coefficient, $\alpha$ & $2.36\times 10^{-5}$\,K$^{-1}$\\
         \hline
         Heat capacity, $c$ & $0.9$\,kJ/(kg\,K) \\
         \hline
         Thermal conductivity, $\lambda$ & $247.0$\,W/(m\,K) \\
         \hline
    \end{tabular}
    \caption{Material data for aluminium from \cite{ozdemir2008fe2}}
    \label{Tab1}
\end{table}

For a better representation of parameters, we use Voigt's notation. In the case of stiffness tensor, the matrix notation reads
\begeq
C\ma_{ijkl} \hat = C\ma_{AB} = 
\begin{pmatrix} 
C\ma_{1111} & C\ma_{1122} & C\ma_{1112} \\
C\ma_{2211} & C\ma_{2222} & C\ma_{2212} \\
C\ma_{1211} & C\ma_{1222} & C\ma_{1212} \\    
\end{pmatrix} \ ,
\eqend
by using $A=\{1,2,3\}$ in place of $ij=\{11,22,12\}$. Analogously, we use $\theta=\{1,2,3,4,5,6\}$ instead of $ijk=\{111, 112, 221, 222, 121, 122\}$ such that we have
\begeq
G\M_{ijklm} \hat = G\M_{A\theta} =
\begin{pmatrix}
G\M_{11111} & G\M_{11112} & G\M_{11221} & G\M_{11222} & G\M_{11121} & G\M_{11122} 
\\
G\M_{22111} & G\M_{22112} & G\M_{22221} & G\M_{22222} & G\M_{22121} & G\M_{22122} 
\\
G\M_{12111} & G\M_{12112} & G\M_{12221} & G\M_{12222} & G\M_{12121} & G\M_{12122} 
\end{pmatrix}
\eqend
as well as
\begeq
D\M_{ijklmn} \hat = D\M_{\theta\gamma} =
\begin{pmatrix}
D\M_{111111} & D\M_{111112} & D\M_{111221} & D\M_{111222} & D\M_{111121} & D\M_{111122} 
 \\
D\M_{112111} & D\M_{112112} & D\M_{112221} & D\M_{112222} & D\M_{112121} & D\M_{112122} 
 \\
D\M_{221111} & D\M_{221112} & D\M_{221221} & D\M_{221222} & D\M_{221121} & D\M_{221122} 
 \\
D\M_{222111} & D\M_{222112} & D\M_{222221} & D\M_{222222} & D\M_{222121} & D\M_{222122} 
 \\
D\M_{121111} & D\M_{121112} & D\M_{121221} & D\M_{121222} & D\M_{121121} & D\M_{121122} 
 \\
D\M_{122111} & D\M_{122112} & D\M_{122221} & D\M_{122222} & D\M_{122121} & D\M_{122122} 
\end{pmatrix} \ ,
\eqend
and
\begeq
\gamma\M_{ijk} \hat = \gamma\M_{Ak} = 
\begin{pmatrix} 
\gamma\M_{111} & \gamma\M_{112} \\
\gamma\M_{221} & \gamma\M_{222} \\
\gamma\M_{121} & \gamma\M_{122} \\    
\end{pmatrix} \ .
\eqend

\subsection{Homogeneous material case}
In homogeneous material case, we expect to retrieve classical continuum mechanics solution where higher order parameters disappear and homogenized values of ${\mathbb{C}\M}$ and ${\mathbb{\beta}\M}$ should maintain isotropic material behavior. This phenomenon is observed in the results below,
\begin{equation*}
C\M_{ijkl} = \begin{pmatrix}
84165.6 & 27774.7 & 0.0 \\ 
27774.7 & 84165.6 & 0.0 \\ 
0.0 & 0.0 & 28195.5\\
\end{pmatrix} \text{MPa}\ , \ 
\beta\M_{ij} = \begin{pmatrix}
-2.64 & 0.0 \\ 
0.0 & -2.64 \\ 
\end{pmatrix} \text{MPa/K} \ ,
\end{equation*}\\
\begin{equation*}
G\M_{ijklm} =
\begin{pmatrix}
0.0 & 0.0 & 0.0& 
0.0 & 0.0 & 0.0 
\\0.0 & 0.0 & 0.0& 
0.0 & 0.0 & 0.0 
\\0.0 & 0.0 & 0.0 & 
0.0 & 0.0 & 0.0 
\end{pmatrix}  \text{N/mm} \ ,
\end{equation*}\\
\begin{equation*}
\gamma\M_{ijk} =
\begin{pmatrix}
0.0 & 0.0 
\\0.0 & 0.0 
\\0.0 & 0.0 
\end{pmatrix}  \text{N/K} \ ,
\end{equation*}\\
\begin{equation*}
D\M_{ijklmn} =
\begin{pmatrix}
0.0 & 0.0 & 0.0 & 
0.0 & 0.0 & 0.0 
 \\
0.0 & 0.0 & 0.0 & 
0.0 & 0.0 & 0.0 
 \\
0.0 & 0.0 & 0.0 & 
0.0 & 0.0 & 0.0 
 \\
0.0 & 0.0 & 0.0 & 
0.0 & 0.0 & 0.0 
 \\
0.0 & 0.0 & 0.0 & 
0.0 & 0.0 & 0.0 
 \\
0.0 & 0.0 & 0.0 & 
0.0 & 0.0 & 0.0 
\end{pmatrix} \text{N} \ ,
\end{equation*}
\begin{equation*}
 \kappa\M_{ij} = \begin{pmatrix}
247.0 & 0.0 \\ 
0.0 & 247.0 \\ 
\end{pmatrix} \text{W/mK} \ , 
\end{equation*}
Heat capacity $c\M$ according to the Eq.\,\eqref{finalMicroParameters} will follow the rule of mixture, while associated parameter $a\M$ will be associated to $c\M$ through the relation in Eq.\,\eqref{relationToHeatCapacity}, as follows:
\begin{equation*}
    c\M = 0.9\, , \,a\M = 0.00225 = \frac{0.9}{400}
\end{equation*}

\subsection{Porous material case}
We emphasize that the higher-gradient is often neglected in composite materials. During the standard homogenization by volume averaging, thermoelastic parameters are obtained in the similar sense as the results herein. Yet the difference, herein, relies on obtaining higher-order parameters as well, where their significance depends on the chosen length-scale, see for a numerical study \cite{abali2022influence}.

\subsubsection{Single central pore case}
In the case of a single centrally located pore, we observe cubic material behavior by inspecting the stiffness tensors components, $C\M_{1111} = C\M_{2222}$. In determining mechanical parameters, we assume constant temperature, $T\M$, over the whole RVE. Thus, thermoelastic interaction, ${\mathbb{\beta}\M}$, is solely a function of geometry. In other words, it affects only the volumetric component of thermal strain. Higher order parameters ${\mathbb{G}\M}$ and ${\mathbb{\gamma}\M}$ are essentially zero due to the centro-symmetry of the RVE with one centrally located pore. Yet higher-order parameters, ${\mathbb{D}\M}$, arise as follows:
\begin{equation*}
C\M_{ijkl} = \begin{pmatrix}
49935.8 & 14164.6 & 0.0 \\ 
14164.6 & 49935.9 & 0.0 \\ 
0.0 & 0.0 & 13570.9 \\ 
\end{pmatrix} \text{MPa} \ , \ 
\beta\M_{ij} = \begin{pmatrix}
-1.51 & 0.0 \\ 
0.0 & -1.51 \\ 
\end{pmatrix} \text{MPa/K} \ ,
\end{equation*}

\begin{equation*}
G\M_{ijklm} =
\begin{pmatrix}
-0.7 & 1.0 & -8.7 & 
-0.7 & -1.2 & 0.0 
\\-0.2 & 0.3 & -2.5 & 
-2.6 & -4.2 & -0.1  
\\-0.7 & -1.1 & 0.0 & 
-0.2 & 0.3 & -2.3& 
\end{pmatrix}  \text{N/mm} \ ,
\end{equation*}

\begin{equation*}
\gamma\M_{ijk} =
\begin{pmatrix}
0.0 & 0.0 
\\0.0 & 0.0 
\\0.0 & 0.0 
\end{pmatrix}  \text{N/K} \ ,
\end{equation*}

\begin{equation*}
D\M_{ijklmn} =
\begin{pmatrix}
-1351.692 & -782.368 & -327.905 & 
-0.001 & -0.002 & 0.003 
 \\
-782.368 & 2006.51 & 626.348 & 
0.0 & 0.001 & 0.0 
 \\
-327.905 & 626.348 & -37.827 & 
0.0 & 0.0 & -0.002 
 \\
-0.001 & 0.0 & 0.0 & 
-1351.697 & -782.372 & -327.905 
 \\
-0.002 & 0.001 & 0.0 & 
-782.372 & 2006.502 & 626.349 
 \\
0.003 & 0.0 & -0.002 & 
-327.905 & 626.349 & -37.816 
\end{pmatrix} \text{N} \ ,
\end{equation*}

\begin{equation*}
 \kappa\M_{ij} = \begin{pmatrix}
205.9 & 0.0 \\ 
0.0 & 205.9 \\ 
\end{pmatrix} \text{W/mK} \ ,   
\end{equation*}
As before, heat capacity $c\M$ according to the Eq.\,\eqref{finalMicroParameters} will follow the rule of mixture (consistent with porosity of 20\%), while associated parameter $a\M$ will be associated to $c\M$ through the relation in Eq.\,\eqref{relationToHeatCapacity}:
\begin{equation*}
    c\M = 0.72 = 0.2 \times 0.9\, , \,a\M = 0.0018 = \frac{0.72}{400}
\end{equation*}

\subsubsection{Four uniformly distributed pores case}
In case of four uniformly distributed pores, we observe the same stiffness tensor, since the porosity is kept the same. We emphasize that the stiffness tensor depends on the porosity; however, the higher-order parameters do depend on the homothetic ratio $\epsilon$. The results for all of the parameters are as follows:
\begin{equation*}
C\M_{ijkl} = \begin{pmatrix}
50009.6 & 14182.9 & 0.0 \\ 
14182.9 & 50009.8 & -0.1 \\ 
0.0 & -0.1 & 13625.3 \\ 
\end{pmatrix} \text{MPa} \ , \
\beta\M_{ij} = \begin{pmatrix}
-1.51 & 0.0 \\ 
0.0 & -1.51 \\ 
\end{pmatrix} \text{MPa/K} \ ,
\end{equation*}

\begin{equation*}
G\M_{ijklm} =
\begin{pmatrix}
0.1 & 1.2 & -3.2 & 
0.2 & -0.5 & -0.8 
\\0.0 & 0.3 & -0.9 & 
0.6 & -1.8 & -2.8 
\\0.1 & -0.5 & -0.7 & 
0.1 & 0.3 & -0.9 
\end{pmatrix} \text{N/mm} \ ,
\end{equation*}

\begin{equation*}
\gamma\M_{ijk} =
\begin{pmatrix}
0.0 & 0.0 
\\0.0 & 0.0 
\\0.0 & 0.0 
\end{pmatrix} \text{N/K} \ ,
\end{equation*}

\begin{equation*}
D\M_{ijklmn} =
\begin{pmatrix}
-338.017 & -194.6 & -81.655 & 
-0.002 & -0.038 & -0.015 
 \\
-194.6 & 500.881 & 155.739 & 
0.026 & 0.001 & 0.003 
 \\
-81.655 & 155.739 & -9.859 & 
0.001 & -0.031 & -0.007 
 \\
-0.002 & 0.026 & 0.001 & 
-337.971 & -194.617 & -81.662 
 \\
-0.038 & 0.001 & -0.031 & 
-194.617 & 500.888 & 155.733 
 \\
-0.015 & 0.003 & -0.007 & 
-81.662 & 155.733 & -9.817 
\end{pmatrix} \text{N} \ ,
\end{equation*}

\begin{equation*}
 \kappa\M_{ij} = \begin{pmatrix}
206.0 & 0.0 \\ 
0.0 & 206.0 \\ 
\end{pmatrix} \text{W/mK} \ ,   
\end{equation*}
For the heat capacity $c\M$ and the associated parameter $a\M$ we can see the same behavior as for the single central circular pore (consistent with porosity of 20\%):
\begin{equation*}
    c\M = 0.72 = 0.2 \times 0.9\, , \,a\M = 0.0017 \approx \frac{0.72}{400}
\end{equation*}

\subsubsection{Four randomly distributed pores case}
In case of four randomly distributed pores, the microscale structure creates an anisotropic material behavior at the macroscale. Thermoelastic interaction, $\mathbb{\beta}\M$, is again a function of geometry so that in the random case, it affects not only the volumetric component of thermal strain but also the shear component of thermal strain. Higher-order parameters ${\mathbb{G}\M}$ and ${\mathbb{\gamma}\M}$ are different from zero as the random distribution circumvents the centro-symmetry of the RVE,  while ${\mathbb{D}\M}$ is also affected by the random distribution of the pores and shows indeed an anisotropic behavior as well. The results for all of the parameters are as follows:
\begin{equation*}
C\M_{ijkl} = \begin{pmatrix}
48494.4 & 15572.7 & 738.1 \\ 
15572.7 & 48305.7 & 124.8 \\ 
738.1 & 124.8 & 14880.7 \\ 
\end{pmatrix} \text{MPa} \ , \
\beta\M_{ij} = \begin{pmatrix}
-1.51 & -0.02 \\ 
-0.02 & -1.51 \\ 
\end{pmatrix} \text{MPa/K} \ ,
\end{equation*}

\begin{equation*}
G\M_{ijklm} =
\begin{pmatrix}
7.8 & -27.2 & 2.6 & 
277.8 & -136.3 & 13.4  
\\-38.6 & 519.1 & 28.9 & 
-131.0 & -340.6 & 19.1  
\\-4.5 & -21.3 & 30.2 & 
-29.0 & 15.5 & -207.0  
\end{pmatrix}  \text{N/mm} \ ,
\end{equation*}

\begin{equation*}
\gamma\M_{ijk} =
\begin{pmatrix}
0.00256 & 0.01498 
\\0.01204 & 0.01823 
\\-0.00324 & 0.00612  
\end{pmatrix}  \text{N/K} \ ,
\end{equation*}

\begin{equation*}
D\M_{ijklmn} =
\begin{pmatrix}
-862.421 & -396.974 & -49.151 & 
-175.055 & 222.629 & 71.895  
 \\
-396.974 & 981.41 & 204.585 & 
201.93 & -154.281 & 49.659  
 \\
-49.151 & 204.585 & -160.335 & 
137.948 & 46.779 & -111.412  
 \\
-175.055 & 201.93 & 137.948 & 
-640.54 & -354.525 & -1.8  
 \\
222.629 & -154.281 & 46.779 & 
-354.525 & 919.989 & 94.141 
 \\
71.895 & 49.659 & -111.412 & 
-1.8 & 94.141 & -273.504 
\end{pmatrix} \text{N} \ ,
\end{equation*}

\begin{equation*}
 \kappa\M_{ij} = \begin{pmatrix}
206.0 & 0.0 \\ 
0.0 & 206.0 \\ 
\end{pmatrix}  \text{W/mK} \ ,   
\end{equation*}
For the heat capacity $c\M$ and the associated parameter $a\M$ we can see the same behavior as for the two previous examples (consistent with porosity of 20\%):
\begin{equation*}
    c\M = 0.72 = 0.2 \times 0.9\, , \,a\M = 0.0017 \approx \frac{0.72}{400}
\end{equation*}
\section{Conclusion}
A higher-order asymptotic homogenization model for generalized thermomechanics has been proposed and implemented by means of strain gradient elasticity and a first-order thermodynamics modeling approach. This model incorporates the effects of microscale morphology through additional (higher-order) thermal and mechanical material parameters at the macroscale. On the mechanical side, we account for stiffness matrix $\mathbb{C}\M$ and higher-order parameters $\mathbb{D}\M$ and $\mathbb{G}\M$, while on the thermal side, we account for thermoelastic interaction $\mathbb{\beta}\M$ and higher-order parameter $\mathbb{\gamma}\M$. All of the thermal and mechanical macroscale parameters are explicitly computed in this work by assuming a linear thermoelastic material model at the microscale.\\

In this framework, the higher-order asymptotic homogenization is implemented in the FEniCS platform and used to solve the partial differential equations generated from the homogenization procedure. The methodology is verified by using a problem without the microstructure leading to homogeneous parameters. Also a porous material is calculated providing insight how the parameters alter for three types of distributions (single, uniform, random distribution). These cases has shown that thermoelastic interaction $\mathbb{\beta}\M$ has a similar reaction to pore morphology as the stiffness matrix $\mathbb{C}\M$ while higher-order thermal parameters $\mathbb{\gamma}\M$ mirror the response of the higher-order mechanical parameter $\mathbb{G}\M$ as both of them are linearly dependent to the size of the RVE and have the same behavior with respect to the centro-symmetry of the RVE. Even though these numerical results for thermal parameters still need in-depth numerical analysis and experimental validation, we provide herein a complete methodology and its implementation to encourage further research for a better understanding of the interplay between microscale morphology and thermo-mechanical material parameters.

\section{Acknowledgment}
This work was supported by a project entitled ``Time-dependent THMC properties and microstructural evolution of damaged rocks in excavation damage zone'' funded by the U.S. Department of Energy (DOE), Office of Nuclear Energy under award \#DE-NE0008771.

\section{Contribution}
\textbf{Bozo Vazic}: Methodology, Software, Validation, Investigation, Writing - Original Draft. \textbf{Bilen Emek Abali}: Methodology, Software, Validation, Writing- Reviewing and Editing. \textbf{Pania Newell}: Conceptualization, Supervision, Funding acquisition, Writing- Reviewing and Editing.

\begin{appendices}
\section{Taylor Expansion of the Logarithm Function}\label{App1}
\begin{figure}[h!]
    \centering
    \includegraphics[scale = 0.6]{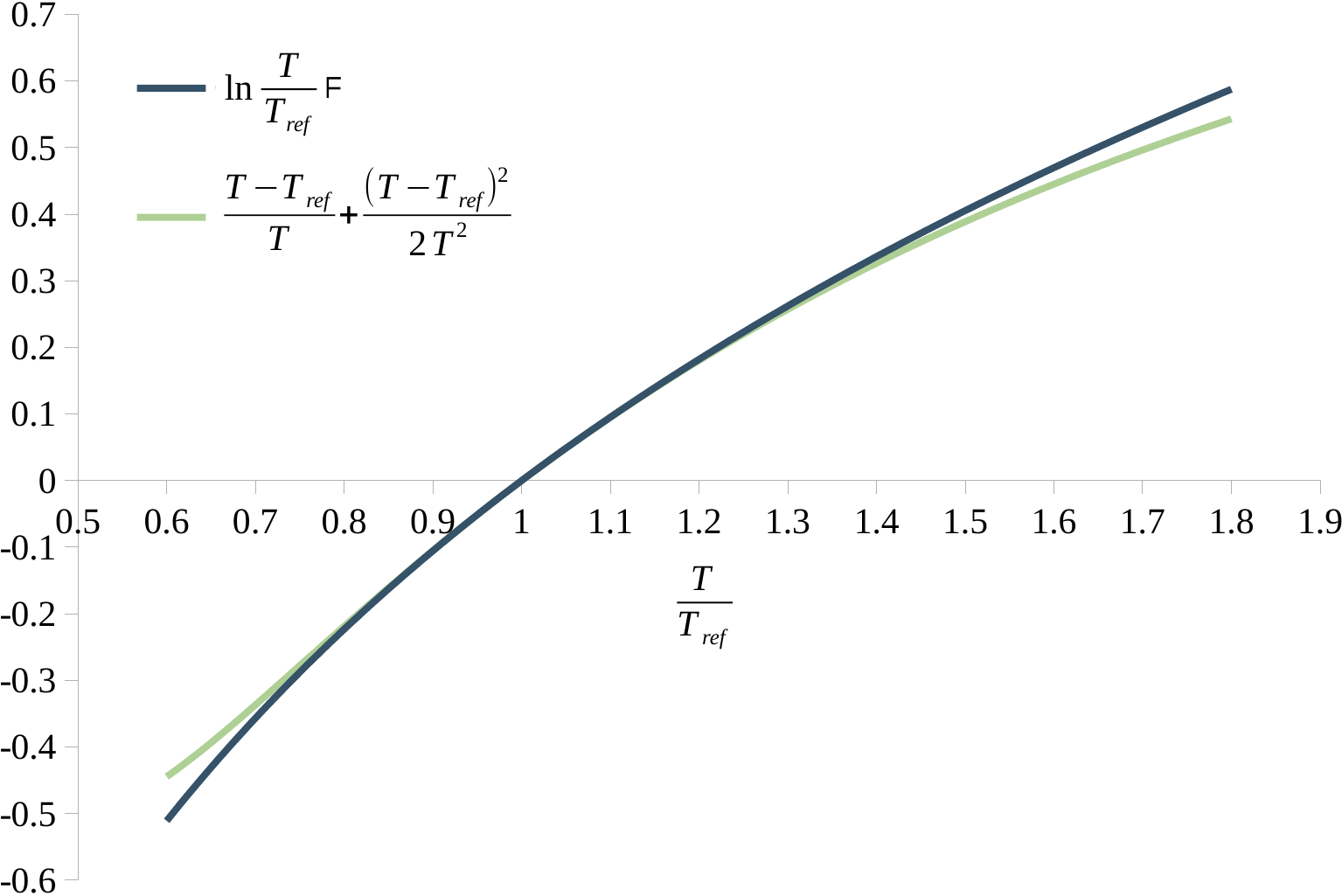}
    \caption{Comparison between $\ln\frac{T}{T_\Reff}$ and Taylor expansion of $\ln\frac{T}{T_\Reff}$}
    \label{tyalorTest}
\end{figure}

First parameter on the left-hand side of Eq.\,\eqref{defEng} needs to be simplified in order to compare microscale and macroscale Helmholtz free energies. This simplification is done by expanding the logarithmic term through Taylor expansion, which may have several different forms depending on the value of $\xi$, see  Eq.\,\eqref{tylor}. What follows is fully developed Eq.\,\ref{tylorExpRHS} for $\xi \geq \frac{1}{2}$, as follows:
\begal\label{tylorExpRHSFull}
\ln \Big(\frac{T\mi}{T_\Reff}\Big) -1 =& \frac{T\mi-T_\Reff}{T\mi} + \frac{(T\mi-T_\Reff)^2}{2(T\mi)^2} - 1 \ , \\
-c\mi T \bigg( \ln\Big(\frac{T}{T_\Reff}\Big)  - 1 \bigg) =&-c\m T\mi\bigg( -\frac{T_\Reff}{T\mi} + \frac{(T\mi-T_\Reff)^2}{2(T\m)^2}\bigg) \\
&-c\m \bigg(-T_\Reff + \frac{(T\mi-T_\Reff)^2}{2T\mi}\bigg) \\
&c\m T_\Reff - \frac{a\m}{2} (T\mi-T_\Reff)^2
\alend
As we are specifying $\xi \geq \frac{1}{2}$, we need to determine the temperature range where Eq.\,\ref{tylorExpRHS} is valid. In Figure \ref{tyalorTest} we show comparisons between $\ln\frac{T}{T_\Reff}$ and its Taylor expansion. If we assume that room temperature, $T_\Reff$, is at $300$\,K, from Figure \ref{tyalorTest}, we see that the expansion is accurate for a temperature range from $180$\,K to $540$\,K.\\

\end{appendices}

\bibliographystyle{special}
\bibliography{Reference}

\end{document}